\documentclass[12pt]{article}
\usepackage{graphicx}
\usepackage{epstopdf}
\usepackage{amssymb,amsmath,epsfig}
\usepackage{cite}
\usepackage[english]{babel}
\usepackage{array}
\usepackage[autostyle]{csquotes}

\begin{document}

\title{\bf Quasinormal Modes, Greybody Factors and Rigorous Bounds for Quantum Oppenheimer-Snyder Black Hole with Quintessential Dark Energy and a String Clouds
}
\author {W. Sajjad \thanks{sajjadwaseem835@gmail.com }, A. Zahid \thanks{asmazahid117@gmail.com}, M. A. Muawia \thanks{bsf2000683@ue.edu.pk} 
and M. Azam \thanks{azam.math@ue.edu.pk, azammath@gmail.com},
\\
Department of Mathematics,\\ Division of Science and Technology,\\ University of Education, Lahore, Pakistan.}
\date{}
\maketitle
\begin{abstract}
We obtained quasinormal modes and greybody factors of the quantum
Oppenheimer-Snyder (QOS) black hole with quintessential dark energy and string clouds. We compute the effective
potentials of the scalar and the vector perturbation fields and analyze their graphical behavior. 
For this, we compute quasinormal mode frequencies of the QOS black hole, that is
surrounded by quintessence dark energy and a cloud of strings and this is done by 6th-order
Wentzel-Kramers-Brillouin (WKB) approximation method. We observe the quasinormal frequencies
by analyzing both the quintessence parameters and the string cloud parameters. We also examine the
GFs by analyzing the role of string cloud ($\tilde{a}_{e}$), quantum deformation ($\tilde{\sigma}_{e}$), and quintessence ($\tilde{h}_{e}$)
parameters. It is found that GFs are significantly
influenced by the parameters of $\tilde{a}_{e}$, $\tilde{\sigma}_{e}$, and $\tilde{h}_{e}$. In the case of scalar perturbation, we also give
strict limits of the GFs and check the influence of the parameters of $\tilde{a}_{e}$, $\tilde{\sigma}_{e}$, and $\tilde{h}_{e}$.
\end{abstract}

\section{Introduction}

Two of the most intriguing predictions of general relativity (GR) are black holes (BHs) and
gravitational waves (GWs). Since decades until today, BHs, one of the most mysterious phenomena
of the universe, have been of interest to the scientific community. The BH information paradox
is one of the most famous gravitational issues. It is believed that more information about
this object will pave the way for the quantum theory of gravity. Much has been created, and
more knowledge has been gained during almost fifty years since this issue was initially
presented, but it still requires a little bit of further consideration. It is possible that
there are very large BHs in a few active galactic nuclei, and the full gravitational collapse
of the material that formed these BHs may have taken place in the past \cite{1*,2*}.  

It is widely believed that once a star large enough undergoes depletion of fuel, the force of gravity
pulling inward takes over and stuff quickly falls inward to the center, resulting in the
formation of a BH. It is crucial to study how a BH reacts to radiation or other matter that
enters it, for instance, in the case of a localized disruption within a BH's motionless outer
field. Since the disruption varies with time, some of the radiation leaks away to infinity, and
some of the radiation is absorbed by the BH. The scattered radiation will reflect the properties
of the BH if the perturbation can ``feel" the BH, that is, if it has relevant components with
wavelengths similar to the size of the BH \cite{3*}. Over the last decade, there have been many
numerical investigations demonstrating that, in late times, the scattered radiation is dominated
by a few damped oscillations.

Although the classical definition of BHs has been quite a success, it is expected to collapse at
the Planck scale, where quantum gravity effects become supreme. In order to overcome these key
limitations, promising quantum gravity treatments have arisen that include quantum corrections
to the equations of Einstein without the need to have a full theory of quantum gravity \cite{4*,5*}. Of
these methods, loop quantum gravity (LQG) has received significant interest because it is a
non-perturbative and background-independent formulation \cite{6*,7*}. Loop quantum gravity is a
discrete spacetime geometry prediction at the Planck scale, which causes classical BH solutions
to become quantum deformed by quantum deformation parameters. Such changes are often seen as new
terms to the metric function that are important in the vicinity of the BH core, which could
eliminate classical singularities and offer a new understanding of BH thermodynamics and stability \cite{8*,9*}.

The QOS model is one such interesting implementation of successful
quantum gravity in the physics of BH \cite{10*}.
The QOS solution gives a regularized description of gravitational collapse
that does not form singularities and retains important BH spacetime properties.
Recent studies have shown that these quantum corrections play important roles
in numerous aspects of BH physics, perturbation dynamics, thermodynamic
properties, and geodesic motion \cite{11*,12*}. In parallel with the
progress in quantum gravity, the cosmic acceleration has led to new
attention to dark energy, and its potential manifestations might occur in BH environments.
The dynamics of quintessence fields (QFs) and BHs have been a subject
of considerable theoretical interest, with these areas able to alter the
geometry of spacetime around BHs and affect their observational behavior
\cite{13*,14*}. 

The other feature of altered BH physics is the topological
defects (especially cosmic strings and their gravitational imprint). Cloud
of strings (CS) configurations are a particular type of matter distribution
capable of enclosing BHs, which is the result of the interplay between
fundamental strings or cosmic strings and the gravitational field \cite{15*,16*}. The
interplay between quantum gravitational effects, QFs, and CS in BH spacetimes
is a largely unexplored theoretical space. Each of the components makes a
specific change to the classical Schwarzschild BH geometry, and their
interaction may give rise to new phenomena that could not occur in the
simplified models. Recent studies have reported various BH solutions
surrounded by CS in combination with additional physical fields. The
influence of Ayon-Beato-Garcia nonlinear electrodynamics with CS was
studied in \cite{17*}, and the BH solution with dark matter (perfect fluid) was introduced.

Another useful method of exploring BH stability and information about
QNMs is perturbation analysis, which encodes characteristic frequencies of BH oscillations \cite{18*,19*}.
These perturbation types, which are scalar (spin-$0$), EM (spin-$1$), and
fermionic (spin-$1/2$) perturbations, probe different aspects of spacetime
geometry and may exhibit evidence of exotic matter fields or quantum gravitational
effects \cite{20*,21*}.  Quasinormal modes are of particular interest in the age of
gravitational wave astronomy, where they can be used to explain the ringdown phase of
BH mergers, and could be able to differentiate between various gravity theories \cite{22*,23*}.

When a BH is perturbed, these damping oscillations can be strongly defined by specific
complex eigenvalues of the wave equations, which are known as quasinormal frequencies.
The real and imaginary components of these frequencies represent the oscillation frequencies
and the damping rates of the modes, respectively. The QNMs of the BH depend only on its
properties, regardless of the manner in which the BH was perturbed. Thus, QNMs are often
referred to as the footprints of a BH. Regge and Wheeler \cite{24*} did pioneering work
on the stability of the BH by looking at the linear perturbation of the Schwarzschild BH. Other
works followed later discussing QNMs and how they contribute to the response of the BH to
external perturbations \cite{25*}. There are clear correlations between the BHs QNMs and
their parameters. The QNMs in the gravitational radiation are used to calculate the BH
parameters because these QNMs have the largest impact on the gravitational radiation
produced due to the vibration of the BH. Consequently, QNMs play a vital role not only
in the search for BHs and their gravitational radiation but also in the study of BH
stability. Numerous studies have focused on BH QNMs within the framework of GR
\cite{26*}. Initial interest arose from the potential to observe quasi-normal
ringing using gravitational wave (GW) detectors  \cite{27*}. 

Currently, QNMs are of significant interest across various contexts,
including the relationship between anti-de Sitter (Ads) and conformal field theory
\cite{28*,29*,30*}, the thermodynamic properties of BHs in LQG \cite{31*}, and their
possible connections with critical collapse \cite{28*,32*}. In recent work, it was
demonstrated that the effective potential of the wave equation simplifies to the
Poschl-Teller potential in the near-extremal limit, especially when studying the
asymptotic values of QNMs of almost extremal Schwarzschild-de Sitter BH. Thus, it
is expected that the Poschl-Teller method provides the most accurate results near
the extremal value of $\Lambda$ \cite{33*}. Furthermore, Leaver \cite{34*} has
generalized the continued fraction methodology to compute the correct values of
the quasinormal frequencies of the ReissnerNordstrom BH, but Kokkotas and Schutz
\cite{35*} have questioned the generalizability of the method to charged BHs. Gonzales
et al. \cite{36*} computed the Hawking radiation and QNMs while investigating a
three-dimensional Godel BH. Moreover, the authors of \cite{37*} have studied the
stability of the quantum-corrected BH by computing the QNMs. The results indicate
that quantum-corrected BH is stable to both scalar and vector perturbations.

A traditional BH absorbs particles but does not expel them. However, quantum mechanics
allows a BH to emit and produce particles beyond its event horizon, and this could be
interpreted as thermal radiation known as Hawking radiation \cite{38*}. This radiation
interacts with the gravitational potential caused by the BH itself, resulting in the
reflection and transmission of the Hawking radiation. As a result, the blackbody spectrum
and the spectrum observed by a distant observer are different. The deviation of Hawking
radiation from a perfect blackbody spectrum is measured by a BH GFs. The study of GFs
offers good information in addition to the analysis of QNMs by defining how the radiation
spectrum changes as it moves out of the gravitational field of a BH. The GFs play a key
role in the energy distribution of radiation emitted by BHs and affect the signals
observed by gravitational wave observatories.

In addition, it was also revealed that the greybody component was significant in the
case of an outgoing wave reaching infinity, or the greybody component significantly
affects the ringdown signal after an excessive mass ratio merger and the absorption
of an incoming wave by a BH \cite{39*,40*,41*,42*}. By integrating the GFs and QNMs, the
BH dynamics, as well as their interaction with the environment, can be completely
understood. The GFs of BHs in the asymptotically de Sitter (dS) spacetime with dRGT
heavy gravity were studied in \cite{43*}, and Hawking radiation through the gray-body
components was computed by Kokkotas et al. \cite{45*} by using the 6th-order WKB technique, and
this technique was proposed by Konoplya \cite{44*}.
Furthermore, the 6th-order WKB method is utilized in many works \cite{46*,47*,48*,48**,48***} in order
to produce Hawking radiation due to its flexibility and versatility.
There are different ways to compute the GFs. The general rigorous bound on the transmission
and reflection coefficients of one-dimensional potential scattering is one such methodology
\cite{49*,50*,51*}. We use the rigorous bound to determine the GFs in this study. The strict
binding gives a qualitative characterization of a BH.
In this paper, we explore the QNMs, GFs, and strict lower limits on the GFs of the QOS BH in
the presence of quintessential dark energy and string clouds. Using this work, we will enhance
our knowledge of BH physics in modified gravity and illuminate the complex interaction between
quantum fields and the surrounding medium of the universe.

The paper is structured in the following way: Section II presents the BH spacetime and discusses
the perturbations and behavior of the effective potential. In Section III, the 6th-order WKB
approximation is applied to the analysis of QNMs. Section IV calculated the frequencies of the
QNMs in the eikonal limit using the circular null method of geodesic. Section V addresses the
GFs by using the 6th-order WKB approximation and the rigorous lower bounds of the GFs.
Lastly, Section VI contains the conclusion and summary of the research.

\section{Introduction to Black Holes and General Formula of Effective Potential}

In the current study, we examine the QOS BH spacetime computed
by Faizuddin Ahmed et al., in ref. \cite{1}, which gracefully includes corrections of loop quantum
gravity together with the influence of quintessential dark energy and a surrounding cloud of
strings, providing a very rich system to investigate exotic gravitational phenomena beyond classical general relativity.
The line element is statical and spherically symmetrical and takes the form
\begin{equation}\label{1}
ds^{2}=-\hat{\bar{S}}(r)dt^{2}+\frac{1}{\hat{\bar{S}}(r)}dr^{2}+r^{2}(d\theta^{2}+sin^{2}\theta d\phi^{2})\\
\end{equation}
and the metric function is written as $\hat{\bar{S}}(r) = 1-\tilde{a}_{e}-\frac{2\check{\bar{M}}}{r}+
\frac{\tilde{\sigma}_{e}\check{\bar{M}}^{2}}{r^{4}}-\frac{\tilde{h}_{e}}{r^{3\hat{\varpi}_{e}+1}} $
with the parameters $\tilde{a}_{e}$, $\tilde{\sigma}_{e}$, and $\tilde{h}_{e}$ represent the contributions
of the string cloud, quantum deformation, and quintessence respectively.
\begin{figure}
    \centering
   {\includegraphics[width=0.95\textwidth]{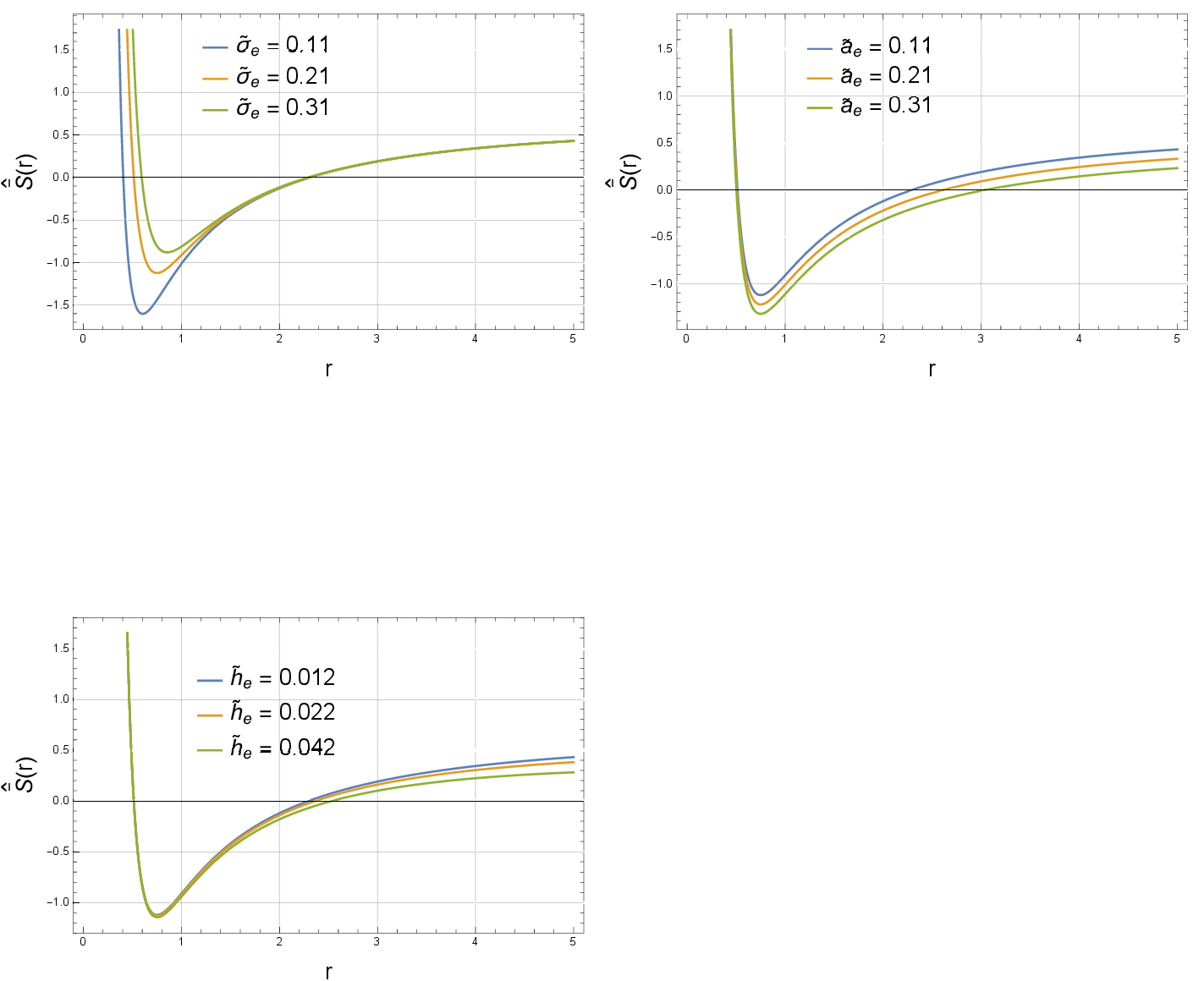}}
%   {\includegraphics[width=0.45\textwidth]{123H22.eps}}
  \caption{The behavior of function $\hat{\bar{S}}(r)$ vs. \( r \) for different
  values of $\tilde{\sigma}_{e}$, $\tilde{a}_{e}$, and $\tilde{h}_{e}$ with fixed values of $\check{\bar{M}}=1$, and $\hat{\varpi}_{e}=-2/3$. }
    \label{fig:multi_graphs}\label{1}
  \end{figure}
Based on the Fig. $\textbf{1}$, it is noticed that by varying different values of parameters, there
are two horizons in each panel. As the quantum correction parameter is increased, the inner horizon
is expanded significantly, and the outer one changes insignificantly; the radius increases overall.
On the other hand, raising the parameter of cloud strings or the quintessence parameter brings two
horizons but leads to a reduction in the metric function and an increase in radius. Here, the dynamics of a massless scalar field
governed by the Klein-Gordon equation, the BH QNMs and GFs of the BH, and the effective potentials
of the scalar and vector perturbation fields are considered. We examine the influence of various
properties inherent to the BH spacetime geometry, such as the parameters, namely, of the
contribution of the string cloud, quantum deformation, and quintessence, i.e., $\tilde{a}_{e}$,
$\tilde{\sigma}_{e}$, and, $\tilde{h}_{e}$, respectively.

The massless scalar and massless vector perturbations of the QOS BH with
quintessential dark energy and with string clouds will be discussed. This background spacetime is
characterized by a massless scalar field $\tilde{\digamma}$ that satisfies the form of the Klein-Gordon equation.
\begin{equation}\label{2}
\Box\tilde{\digamma}\equiv \frac{1}{\sqrt{-\check{\bar{g}}}}\partial_{\hat{\rho}}\Big(\sqrt{-\check{\bar{g}}}
\check{\bar{g}}^{\hat{\rho}\hat{\delta}}\partial_{\hat{\delta}}\tilde{\digamma}\Big)=0,
\end{equation}
and the above equation can be solved by making sure.\\
\begin{equation}\label{3}
\tilde{\digamma}(t, r, \theta, \phi)=\frac{\check{\tilde{e}}^{-i \check{\varpi} t}}{r}\breve{\digamma}(r)\bar{E}^{m}_{l}(\theta,\phi),
\end{equation}
and where the frequency of the wave $\tilde{\digamma}$ is denoted by $\check{\varpi}$, and
the wave-function in the radial direction is indicated by $\breve{\digamma}(r)$.
Replacing the Eq. $(\ref{3})$ in the Eq. $(\ref{2})$ the Regge-Wheeler \cite{2}
wave equation can be obtained and it can be written as
\begin{equation}\label{4}
\frac{d^{2}}{dr^{2}_{*}}\breve{\digamma}+(\check{\varpi}^{2}-\bar{\check{V}}_{eff})\breve{\digamma}=0,
\end{equation}
apply to the effective potential which is denoted by the symbol $\bar{\check{V}}_{eff}$ and the radial
coordinate (so-called tortoise) is denoted by $r_{*}=\int\frac{dr}{\hat{\bar{S}}(r)}$.
\begin{equation}\label{5}
\bar{\check{V}}_{eff}(r)=\hat{\bar{S}}(r) \Big[\frac{l_{y}(l_{y}+1)}{r^{2}}+\frac{1}{r}\frac{d\hat{\bar{S}}(r)}{dr}\Big].
\end{equation}
The proper potential of higher spin (boson) fields is obtainable through the generalised equation \cite{3,4}:
\begin{equation}\label{6}
\bar{\check{V}}_{eff}(r)=\hat{\bar{S}}(r) \Big[\frac{l_{y}(l_{y}+1)}{r^{2}}+(\check{z}-\check{z}^{2})\frac{1-
\hat{\bar{S}}(r)}{r^{2}}+(1-\check{z})\frac{1}{r}\frac{d\hat{\bar{S}}(r)}{dr}\Big],
\end{equation}
here, the spin of the perturbative field is indicated by a $\check{z}$,
\begin{center}
$
\check{z}=
\begin{cases}
0, & \text{scalar perturbation,}\\
1, & \text{electromagnetic perturbation, }  \\
2, & \text{gravitational perturbation, }  \\
\end{cases}
$
\end{center}
where $\check{z}\leq l_{y}$.\\
The effective potential of the scalar perturbation of $\bar{\check{V}}_{sc}(r)$ and the
gravitational perturbation of $\bar{\check{V}}_{gr}(r)$ are not equal to the vector perturbation
$\bar{\check{V}}_{em}(r)$ because scalars and gravitational perturbations include derivative terms, and vectors do not.
In the study of QOS BH perturbations, an effective potential plays a
critical role, particularly in the study of QNMs, as well as the GFs in both scenarios. In perturbation
theory, a radial equation of a massless field, electromagnetic field, or gravitational field near a
BH can be simplified to a Schrodinger-like equation, with the effective potential acting as a
barrier that decides whether or not the waves propagate and or scatter. 

In the case of QNMs, the
complex frequencies of the BH in the relaxation phase after a perturbation by the effective potential
can be used to determine the complex eigenfrequencies. The complex eigenfrequencies are determined
by applying ingoing and outgoing boundary conditions to the waves on the event horizon and at infinity
and, thereby, display the stability, phase transitions, and thermodynamics.
They are GFs: the differences between Hawking radiation of a BH and perfect blackbody
emission are due to the effective potential of the BH, which serves to block them. This possible state of
spacetime, which is produced by the curvature of spacetime and the angular momentum, blocks
the escape of particles and waves by tunneling or propagation.
The more intense and wider this barrier, the greater its effect in blocking low-frequency
radiation, producing a greybody spectrum that is much weaker and more filtered than an ideal
blackbody, and this has an impact on the total rate of evaporation of the BH.\\

\subsection{Effective Potential}

Here, we provide a short summary of the characteristics of the perturbation potentials of the
BH under consideration. The possible behavior has been explored to gain general knowledge of
the QNMs and their dependence on the different parameters.
\begin{figure}[h!]
    \centering
   {\includegraphics[width=0.95\textwidth]{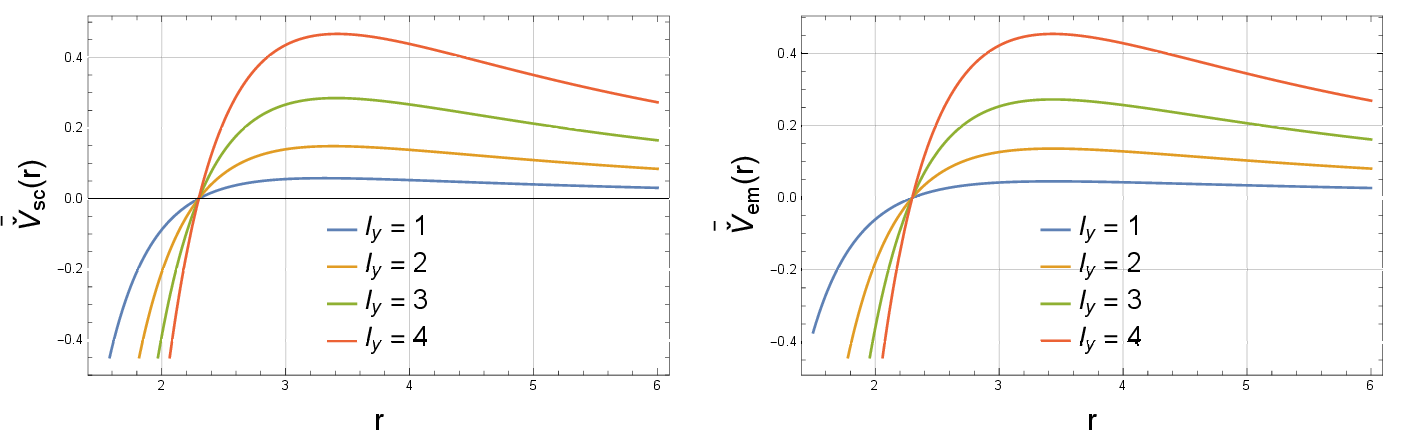}}\hfill
    \caption{The scalar potential $\bar{\check{V}}_{sc}(r)$ (left panel) and the electromagnetic potential $\bar{\check{V}}_{em}(r)$ (right panel) versus $r$ for different multipole moment $l_{y}$ values with fixed values of $\check{\bar{M}} = 1$, $\tilde{a}_{e} = 0.11$, $\tilde{\sigma}_{e} = 0.21 $, $\tilde{h}_{e} = 0.012$, and $\hat{\varpi}_{e}=-2/3$.}
    \label{fig:multi_graphs}\label{FIG.2}
\end{figure}
We observe that in Fig. $\textbf{2}$, as the multipole moment $l_{y}$ increases, it results in an
increased effective potential barrier. This causes the BH to vibrate at higher frequencies to form
QNMs; in other words, the vibrations of the BH occur at a faster rate, such as the vibration of a
tighter or stiffer string when plucked, which produces a higher pitch of the sound. Meanwhile, the
damping rate increases, and these oscillations will decay more rapidly due to the increased barrier,
which is able to leak perturbation energy faster through wave emission or scattering. In general,
the modes with greater multipoles are vibrated at faster rates but shorter periods, and are not as
evident in the gravitational wave signal as the lower modes are. In the case of GFs, which quantify
relative absorption and reflection of radiation by the BH, they tend to decrease the GFs of lower-energy
waves when the potential barrier is larger and broader, and, consequently, more radiation is reflected
and less is emitted on the other side of the barrier.
\begin{figure}[h!]
    \centering
   {\includegraphics[width=0.95\textwidth]{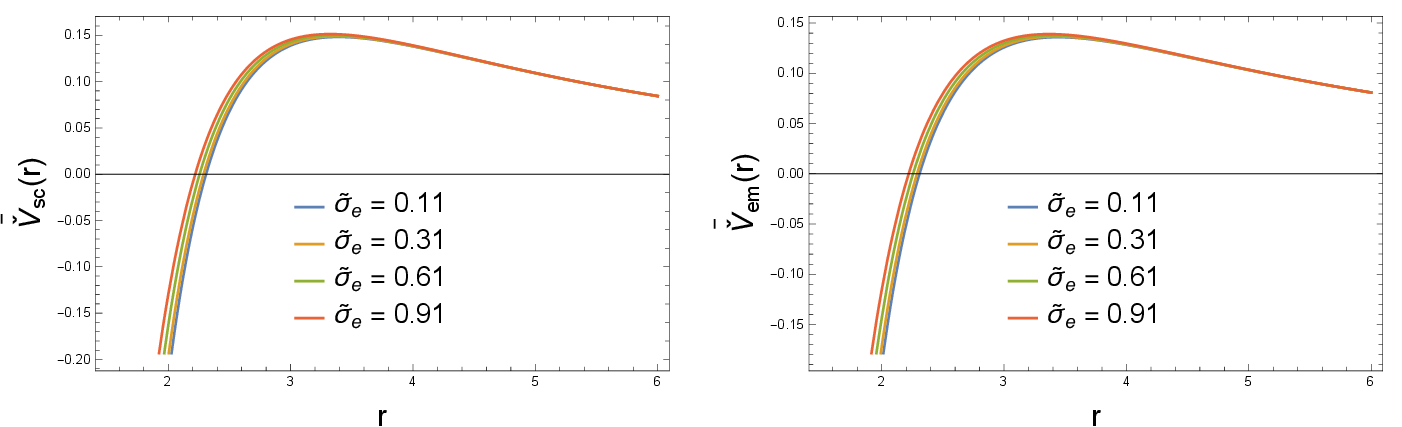}}\hfill
    \caption{The scalar potential $\bar{\check{V}}_{sc}(r)$ (left panel) and the electromagnetic potential $\bar{\check{V}}_{em}(r)$ (right panel) versus $r$ for different quantum correction parameter $\tilde{\sigma}_{e}$ values with fixed values of $\check{\bar{M}} = 1$, $\tilde{a}_{e} = 0.11$, $\tilde{h}_{e} = 0.012$, $l_{y} = 2$ and, $\hat{\varpi}_{e}=-2/3$.}
    \label{fig:multi_graphs}\label{FIG.3}
\end{figure}

Figure $\textbf{3}$ illustrates that as the quantum correction parameter
$\tilde{\sigma}_{e}$ increases, the height and width of the effective potential
barrier increase. This causes QNMs, which are vibrating at higher frequencies,
producing faster vibrations with a higher pitch, and the damping rates are also
higher, making the vibrations decay more quickly since perturbation energy can
escape more easily with the increased barrier. This means that these modes are
resonant with shorter lifetimes and hence even less detectable in gravitational
wave signals than the lower and more dominant modes. At the same time, the higher
and wider barrier inhibits the GFs, especially of low-energy particles, which
decreases the probability of transmission, lowers the total rate of Hawking
radiation emission, causes a shift in the spectrum to lower intensity or energy,
and diminishes the radiative output of the BH in comparison to the classical Schwarzschild scenario.
\begin{figure}[h!]
    \centering
   {\includegraphics[width=0.95\textwidth]{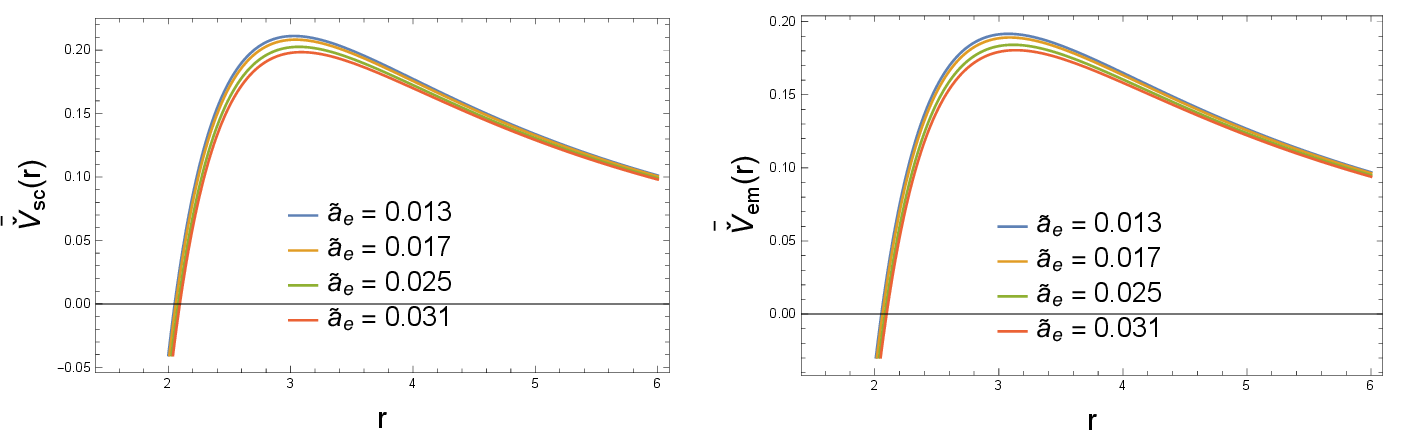}}\hfill
    \caption{The scalar potential $\bar{\check{V}}_{sc}(r)$ (left panel) and the electromagnetic potential $\bar{\check{V}}_{em}(r)$ (right panel) vs. $r$ for various string parameter $\tilde{a}_{e}$ values with fixed values of $\check{\bar{M}} = 1$, $\tilde{\sigma}_{e}= 0.21$, $\tilde{h}_{e} = 0.012$, $l_{y} = 2$ and, $\hat{\varpi}_{e}=-2/3$.}
    \label{fig:multi_graphs}\label{FIG.4}
\end{figure}
\begin{figure}[h!]
    \centering
   {\includegraphics[width=0.95\textwidth]{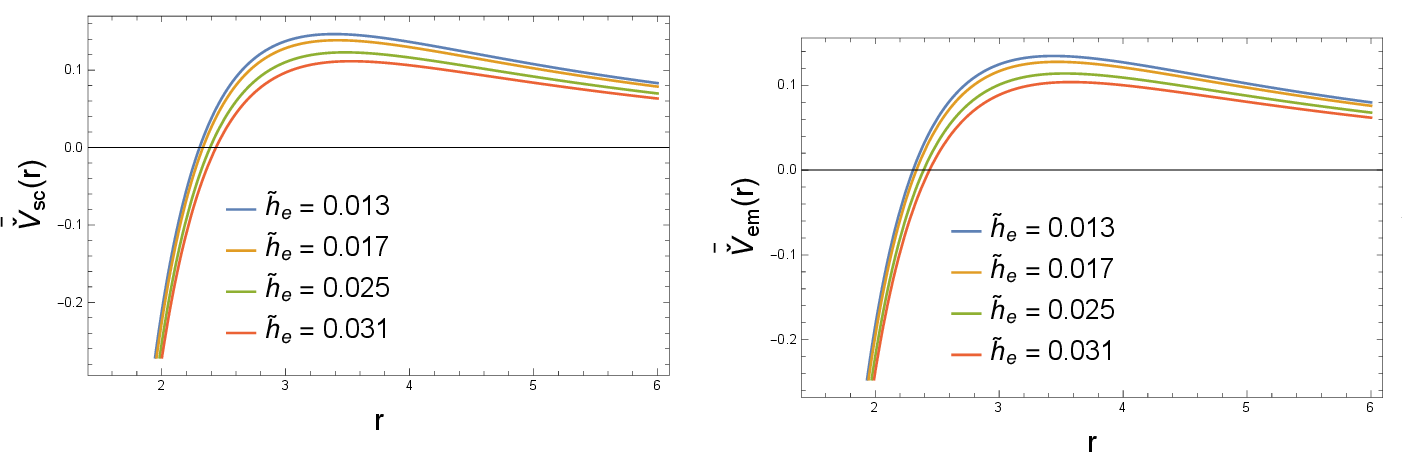}}\hfill
    \caption{The scalar potential $\bar{\check{V}}_{sc}(r)$ (left panel) and the electromagnetic potential $\bar{\check{V}}_{em}(r)$ (right panel) vs. $r$ for various quintessence state parameter $\tilde{h}_{e} $ values, where, $\check{\bar{M}} = 1$, $\tilde{\sigma}_{e}= 0.21$, $\tilde{a}_{e} = 0.11$, $l_{y} = 2$ and, $\hat{\varpi}_{e}=-2/3$.}
    \label{fig:multi_graphs}\label{FIG.5}
\end{figure}
Figures $\textbf{4}$ and $\textbf{5}$ demonstrate that as the cloud of strings parameter
$\tilde{a}_{e}$ and the quintessence parameter $\tilde{h}_{e} $ are increased, the potential
barrier to the effective potential decreases. This decreases the barrier height and results
in lower oscillation frequency QNMs; i.e, the BH vibrates more slowly, like a looser
string that takes a deeper, lower pitch of the sound. The damping rates also reduced at
the same time, and thus these vibrations are slower to decay since the weaker barrier
prevents them to a greater extent from losing the perturbation energy in a short time
through gravitational waves or scattering. 

Consequently, the modes resonate at lower
frequencies but have longer-lived existence periods, and this may mean that higher
multipole modes will be visible or detectable for longer periods in gravitational-wave
signals than in the case when stronger barriers are present. In the case of the GFs, the
lower barrier enhances the probability of transmission, particularly that of low-energy
particles and waves, which enhances the GFs and increases the tunneling of the barrier,
as well as generically increases the overall rate of Hawking radiation emission and tends
to produce a stronger or more intense radiation spectrum. Essentially, both the string
clouds and the quintessence make the geometry of the spacetime surrounding the BH
smoother, leading to slower and longer vibrations as well as increased leakage of Hawking radiation.

The plots of both scalar and vector perturbations, as observed in Figs. $\textbf{2}$,
$\textbf{3}$, $\textbf{4}$, and $\textbf{5}$, qualitatively resemble each other; i.e,
there are similarities in the actions of such parameters as multipole moment $l_{y}$,
quantum correction parameter $\tilde{\sigma}_{e}$, dependence on the cloud of strings
$\tilde{a}_{e}$, dependence on the quintessence $\tilde{h}_{e}$, and the fact that they
are similar in both types of perturbations. However, a primary difference is noticeable: the
scalar perturbation always shows higher potential barrier heights than the vector perturbation. This
means that the scalar field will have a greater interaction with the BH spacetime; the
more barriers, the more the interaction.

Consequently, we expect that the lower oscillation frequencies (smaller real part) and
lower damping rate of the spectral QNMs will be observed as the cloud of strings and
quintessence parameters increases, and then the effective potential barrier decreases. The
frequencies obtained in the scalar perturbation case are still greater than the frequencies
in the case of the vector perturbation, but the oscillation frequencies and damping period
are lower in both cases, and the modes decay more slowly. Such a potential analysis in both
scalar and vector perturbations carries significant information on QNMs properties and
dependence on the above-mentioned parameters, which can be more accurately used to
predict BH stability and dynamical behavior in modified gravity or in the presence
of a surrounding field.

\section{Analysis of Quasinormal Mode with the WKB Method}

We apply the WKB methodology, a popular approximation method in BH perturbation theory, in
this section to estimate the QNMs of BHs. Schutz and Will introduced the WKB method for
computing QNMs, and their first-order approximation of the wave equation is presented in
Ref. \cite{5}. Although it is a valuable method, it is known to have quite a number of
disadvantages, including an increased level of error in certain situations. Researchers
have formulated higher-order WKB approximations to address these limitations, which have
greatly improved the accuracy of computations of QNM. 

Consequently, the ringing stage of the
QNMs coincides with the QNM of the BH, which frequently has damped oscillations with discrete
complex frequencies, in which the frequency of the oscillation is referred to as the real part
and the decay rate as the imaginary part. QNMs are crucial in the exploration of BHs and gravity
theories since they are not dependent on the initial perturbation and are solely dependent on the BH parameters.
There are several methods to determine the frequencies of the
QNMs. In order to enhance the effectiveness of the computation of these frequencies, we
will utilize the 6th-order WKB approximation, which was developed in Ref. \cite{6}, and
we will analyze the scattering around BHs. We consider the basic mode of having $l_{y} = 2$
and $\breve{\varsigma}= (0, 1$), which is primarily influenced by the gravitational waves. We
shall analyze the effects of the quintessence parameter and the string cloud parameter on the quasinormal frequencies.
The 6th-order WKB approximation complex frequency formula is as follows.
\begin{equation}\label{7}
i\frac{(\check{\varpi}^{2}-\bar{\check{V}}_{0})}{\sqrt{-2\bar{\check{V}}^{\prime\prime}_{0}}}-\sum_{b=2}^{6}\breve{\Psi}_{b}= \breve{\varsigma}+\frac{1}{2}
\end{equation}
The maximum value of the effective potential in the above equation is
$\bar{\check{V}}_{0}$, $\bar{\check{V}}^{\prime\prime}_{0}=\frac{d^{2}\bar{\check{V}}(r_{*})}{dr^{2}_{*}}\Big|_{r_{*}=\textit{r}_{\circ}}$, $\textit{r}_{\circ}$
is highest value of the 2nd-order derivative of the effective potential, and Refs.\cite{6,7,8,9} show that the error terms of the 6th order WKB are $\breve{\Psi}_{b}^,s$.

\subsection{QNMs of QOS Black Hole with Quintessential Dark Energy and a String Clouds}

In this subsection, the QNM frequencies of the respective line element
are calculated by the 6th WKB method. Tables $1$ and $2$ show the results of scalar and
electromagnetic perturbations. These tables give the complex QNM frequencies, calculated
at different values of the quintessence parameter $\tilde{h}_{e}$ and the cloud strings
parameter $\tilde{a}_{e}$ with fixed values of mass $\check{\bar{M}}=1$ and multipole moments
$l_{y}=2$. 

The oscillation frequency of the ringing modes is the real part of the QNM
frequency, i.e., the rate at which the BH vibrates when perturbed. As the imaginary
component is negative in both cases under consideration, the perturbations exponentially
decrease over time, which shows that the BH is dynamically stable to a scalar and electromagnetic
perturbation. These results are in agreement with the fact that the existence of quintessence and
the cloud-of-strings parameter typically change the oscillation frequency as well as the damping
timescale. The impact of the model parameters on the QNM spectrum is depicted in both the real (Re)
and imaginary (Im) values of the QNMs as functions of the model parameters.
\begin{table}[h!]
\caption{\label{tab1}
The BH QNMs of scalar and electromagnetic perturbations at specified values of $\tilde{a}_{e}$. Here,
the values of $\check{\bar{M}} = 1$,~$l_{y} = 2$, $\tilde{\sigma}_{e} = 0.5$ and $\tilde{h}_{e} = 0.03$.}
    \centering
    \small % Table ko chhota karne ke liye
    \renewcommand{\arraystretch}{1.2} % Row height ko adjust karne ke liye
    \setlength{\tabcolsep}{9pt} % Column width ko adjust karne ke liye
    \resizebox{1\textwidth}{55pt}{ % Puri table ki width kam karne ke liye
    \begin{tabular}{c|cc|cc}
    \hline
     & \multicolumn{2}{c|}{\textbf{Scalar perturbation}} & \multicolumn{2}{c}{\textbf{Electromagnetic perturbation}} \\
    \hline
    \textbf{$\tilde{a}_{e}$} & \textbf{$\breve{\varsigma}=0$} & \textbf{$\breve{\varsigma}=1$} & \textbf{$\breve{\varsigma}=0$} & \textbf{$\breve{\varsigma}=1$} \\
    \hline
    $0.013$  & $0.4168-0.0574i$  & $0.4207-0.1202i$ &  $0.3983-0.0554i$ & $0.3833-0.1057i$ \\

    $0.015$  & $0.4149-0.0570i$  & $0.4167-0.1184i$ &  $0.3965-0.0551i$ & $0.3765-0.1037i$ \\

    $0.017$  & $0.4130-0.0567i$  & $0.4124-0.1166i$ &  $0.3946-0.0548i$ & $0.3689-0.1016i$ \\

    $0.019$  & $0.4111-0.0564i$  & $0.4077-0.1149i$ &  $0.3927-0.0545i$ & $0.3602-0.0995i$ \\

    $0.021$  & $0.4091-0.0561i$  & $0.4025-0.1131i$ &  $0.3908-0.0543i$ & $0.3501-0.0973i$  \\

    $0.023$  & $0.4072-0.0558i$  & $0.3968-0.1114i$ &  $0.3889-0.0540i$ & $0.3384-0.0949i$ \\

    $0.025$  & $0.4052-0.0555i$  & $0.3904-0.1096i$ &  $0.3870-0.0538i$ & $0.3245-0.0924i$ \\

    $0.027$  & $0.4032-0.0553i$  & $0.3831-0.1078i$ &  $0.3851-0.0536i$ & $0.3078-0.0897i$      \\

    $0.029$  & $0.4012-0.0550i$  & $0.3748-0.1059i$ &  $0.3831-0.0533i$ & $0.2875-0.0869i$      \\

    $0.031$  & $0.3992-0.0548i$  & $0.3651-0.1040i$ &  $0.3810-0.0531i$ & $0.2621-0.0838i$      \\

    \hline
    \end{tabular}
    }
\end{table}

\begin{figure}[h!]
    \centering
   {\includegraphics[width=0.95\textwidth]{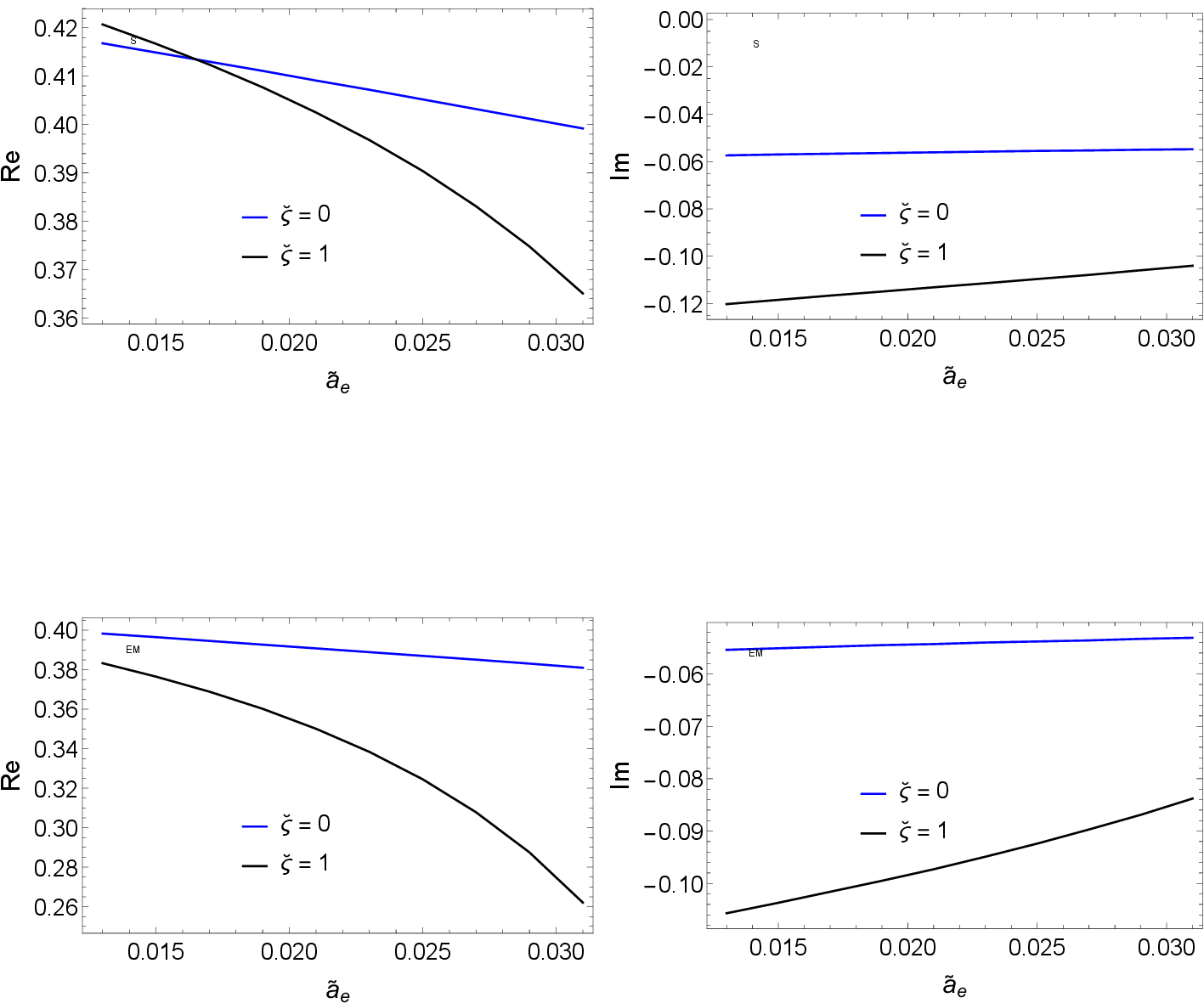}}\hfill

    \caption{Scalor (upper panel) and electromagnetic (lower panel) quasinormal mode frequencies
    versus the string cloud parameter $\tilde{a}_{e}$ with a fixed parameters are  $\check{\bar{M}} = 1$,
    $l_{y} = 2$, $\tilde{\sigma}_{e} = 0.5$ and, $\tilde{h}_{e} = 0.03$.}
    \label{fig:multi_graphs}\label{FIG.6}
\end{figure}

Figure $\textbf{6}$ shows that the real part and the absolute value of the imaginary part of the QNM
frequency decrease with an increase in cloud string parameter $\tilde{a}_{e}$ in the case
of scalar and electromagnetic perturbation. This implies that the rate of oscillation is
reduced, the perturbations vibrate more slowly, and it requires a longer period to complete
a cycle. Simultaneously, the damping rate is also weaker the magnitude of the perturbation
decays more slowly, and the mode decays more slowly; thus, the ringdown process is slower,
and the ringdown mode dies out slowly. In general, the bigger the cloud string parameter, the
less intense the BH response to perturbations, the slower the oscillation, the slower the
energy loss, and the slower the system will jump to equilibrium.
\begin{table}[h!]
\caption{\label{tab1}
The BH QNMs of scalar and electromagnetic perturbations at specified values of $\tilde{h}_{e}$. Here,
the values of $\check{\bar{M}} = 1$,~$l_{y} = 2$, $\tilde{\sigma}_{e} = 0.5$ and $\tilde{a}_{e} = 0.03$.}
    \centering
    \small % Table ko chhota karne ke liye
    \renewcommand{\arraystretch}{1.2} % Row height ko adjust karne ke liye
    \setlength{\tabcolsep}{9pt} % Column width ko adjust karne ke liye
    \resizebox{1\textwidth}{55pt}{ % Puri table ki width kam karne ke liye
    \begin{tabular}{c|cc|cc}
    \hline
     & \multicolumn{2}{c|}{\textbf{Scalar perturbation}} & \multicolumn{2}{c}{\textbf{Electromagnetic perturbation}} \\
    \hline
    \textbf{$\tilde{h}_{e}$} & \textbf{$\breve{\varsigma}=0$} & \textbf{$\breve{\varsigma}=1$} & \textbf{$\breve{\varsigma}=0$} & \textbf{$\breve{\varsigma}=1$} \\
    \hline
    $0.013$  & $0.4483-0.0626i$  & $0.4604-0.1369i$ &  $0.4269-0.0602i$ & $0.4280-0.1216i$ \\

    $0.015$  & $0.4429-0.0616i$  & $0.4529-0.1328i$ &  $0.4219-0.0593i$ & $0.4190-0.1175i$ \\

    $0.017$  & $0.4375-0.0606i$  & $0.4450-0.1288i$ &  $0.4169-0.0584i$ & $0.4091-0.1136i$ \\

    $0.019$  & $0.4319-0.0597i$  & $0.4367-0.1250i$ &  $0.4117-0.0575i$ & $0.3980-0.1097i$ \\

    $0.021$  & $0.4263-0.0587i$  & $0.4277-0.1213i$ &  $0.4066-0.0567i$ & $0.3851-0.1058i$  \\

    $0.023$  & $0.4207-0.0578i$  & $0.4179-0.1177i$ &  $0.4013-0.0559i$ & $0.3699-0.1018i$ \\

    $0.025$  & $0.4149-0.0570i$  & $0.4068-0.1141i$ &  $0.3959-0.0551i$ & $0.3513-0.0976i$ \\

    $0.027$  & $0.4091-0.0561i$  & $0.3941-0.1106i$ &  $0.3905-0.0543i$ & $0.3275-0.0931i$  \\

    $0.029$  & $0.4032-0.0553i$  & $0.3789-0.1069i$ &  $0.3849-0.0536i$ & $0.2959-0.0881i$   \\

    $0.031$  & $0.3972-0.0545i$  & $0.3602-0.1029i$ &  $0.3792-0.0528i$ & $0.2507-0.0826i$  \\

    \hline
    \end{tabular}
    }
\end{table}

\begin{figure}[h!]
    \centering
   {\includegraphics[width=0.95\textwidth]{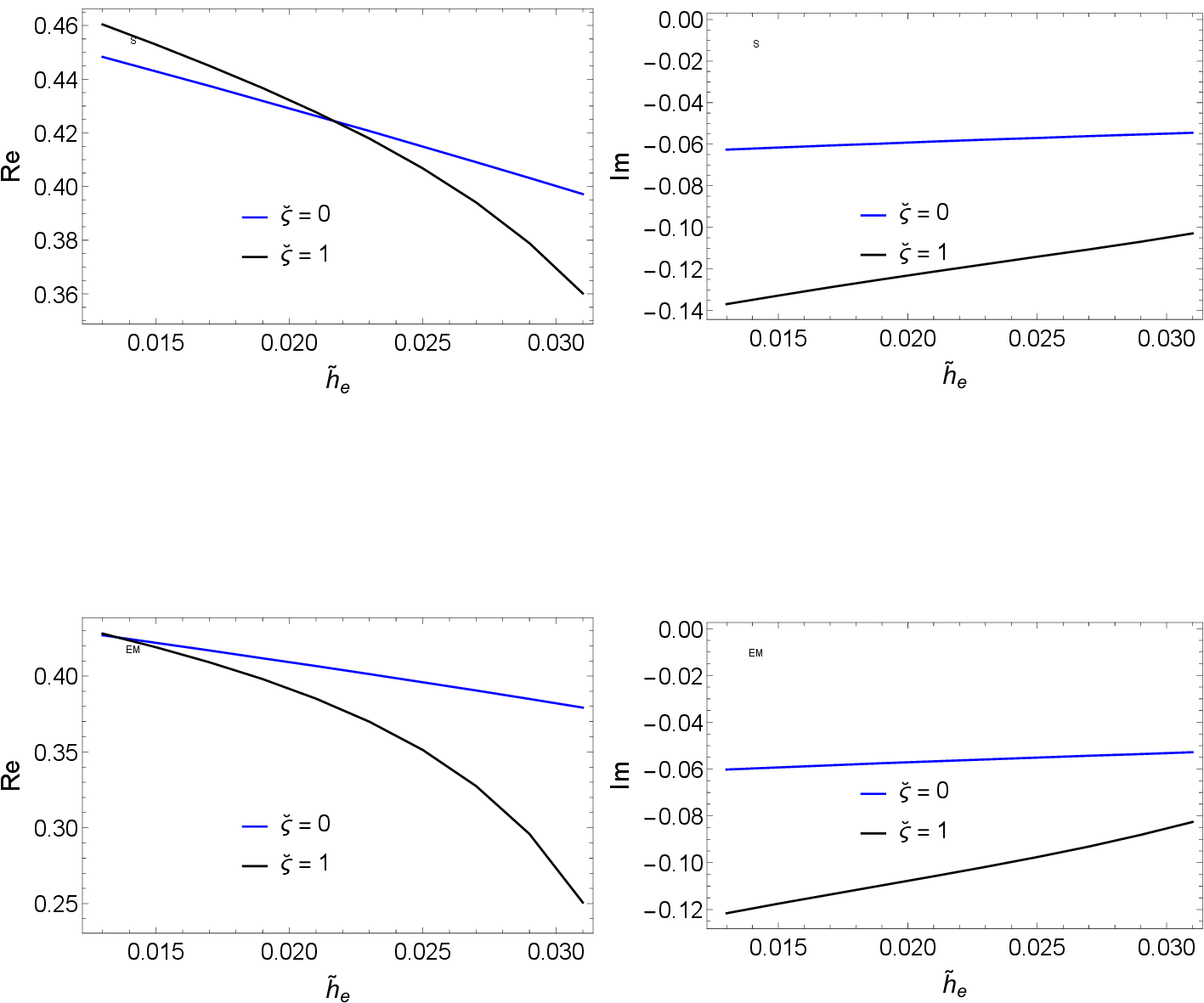}}\hfill
    \caption{Scalor (upper panel) and electromagnetic (lower panel) quasinormal mode frequencies
    vs. quintessence parameter $\tilde{h}_{e}$ with a fixed parameters are $\check{\bar{M}} = 1$,
    $l_{y} = 2$, $\tilde{\sigma}_{e} = 0.5$, and, $\tilde{a}_{e} = 0.03$.}
    \label{fig:multi_graphs}\label{FIG.7}
\end{figure}

Figure $\textbf{7}$ demonstrates that as the quintessence parameter $\tilde{h}_{e}$ increases, both the QNM
frequencies of the scalar and electromagnetic perturbations exhibit a gradual and steady decrease
in their real part and the absolute value of their imaginary part. Consequently, the whole
ringdown process becomes more relaxed and more stretched out: the vibrations become slower, the
energy decays at a slower rate, and the BH takes significantly longer to regain its
silent equilibrium state. The curvature of the spacetime of the BH is softened by
quintessence; thus, the oscillations slow, the damping is weaker, and therefore the quasinormal
signature is less intense and longer.

The Figs. $\textbf{6}$ and $\textbf{7}$ provide a consistent physical picture: the larger either the string cloud or
quintessence parameter, the lower the oscillation frequency and sweptness of damping, 
and longer the quasinormal ringing of scalar and electromagnetic modes. It means that these exotic
fields (string clouds or quintessence) provide a lowering of the effective spacetime potential in
the area of the BH.

\section{The Frequencies of the QNM Calculated by Circular Null Geodesics in the Eikonal Limit}

Cardoso et al. \cite{10NullQNMs} introduced the circular null geodesic method to calculate the
QNM frequencies of a static spherically symmetric BH at the eikonal limit where $l_{y}\gg1$.
For comparison of the WKB method, we also compute the frequencies of the QNM in the eikonal
(high-frequency) limit of the unstable circular null geodesics (photon sphere) as given in
 \cite{10NullQNMs}:
\begin{equation}\label{8}
\check{\varpi} = l_{y}\underline{\hat{\psi}}-i\bigg(\breve{\varsigma}+\frac{1}{2}\bigg)|\underline{\hat{E}}|.
\end{equation}
The real and imaginary values of the QNM frequencies are defined as the angular velocity $\underline{\hat{\psi}}$
and Lyapunov exponent $\underline{\hat{E}}$, respectively. The angular velocity and the Lyapunov exponent are as follows
\begin{equation}\label{9}
\underline{\hat{\psi}}=\frac{\sqrt{\hat{\bar{S}}(r_{q})}}{r_{q}},~~~~~~~\underline{\hat{E}}=
\frac{1}{\sqrt{2}r_{q}}\sqrt{\hat{\bar{S}}(r_{q})[2\hat{\bar{S}}(r_{q})-r^{2}_{q}\hat{\bar{S}}^{\prime\prime}(r_{q})]},
\end{equation}
and $r_{q}$ is the circular path of the unstable null geodesics, which is determined as follows
\begin{equation}\label{10}
2\hat{\bar{S}}(r)-r^{2}\hat{\bar{S}}^{\prime\prime}(r)\bigg|_{r=r_{q}}=0.
\end{equation}
Angular frequency and Lyapunov exponent are computed and illustrated in Figs. $\textbf{8}$ and $\textbf{9}$
respectively as a function of both $\tilde{a}_{e}$ and $\tilde{h}_{e}$. These graphs are obtained
by substituting the metric function $\hat{\bar{S}}(r)$ with the standard forms of the functions
$\underline{\hat{\psi}}$ and $\underline{\hat{E}}$ in Eq. $(\ref{9})$. We observed that the
qualitative and quantitative behaviors of Figs. $\textbf{8}$ and $\textbf{9}$ are similar to
Figs. $\textbf{6}$ and $\textbf{7}$ of the QNM frequencies. This is a very robust agreement and
so validates that the higher-order WKB approximation and the geometric method that was derived
by Cardoso et al. \cite{10NullQNMs} (in terms of the characteristics of unstable null geodesics
in the eikonal limit) are essentially the same in their results on the QNM frequencies
of this spacetime and so validates the consistency and accuracy of each method.
\begin{figure}[h!]
    \centering
   {\includegraphics[width=0.95\textwidth]{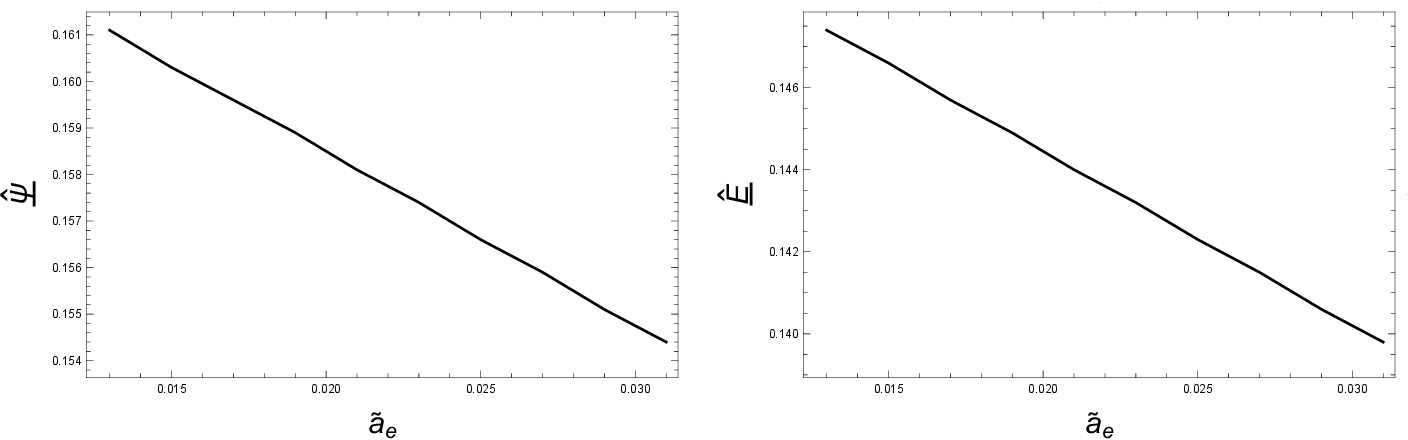}}\hfill
    \caption{Angular velocity $\underline{\hat{\psi}}$ (left panel) and Lyapunov exponent $\underline{\hat{E}}$ (right panel)
    versus the string cloud parameter $\tilde{a}_{e}$ with a fixed parameters are  $\check{\bar{M}} = 1$,
    $l_{y} = 2$, $\tilde{\sigma}_{e} = 0.5$, and, $\tilde{h}_{e} = 0.03$.}
    \label{fig:multi_graphs}\label{FIG.8}
\end{figure}
\begin{figure}[h!]
    \centering
   {\includegraphics[width=0.95\textwidth]{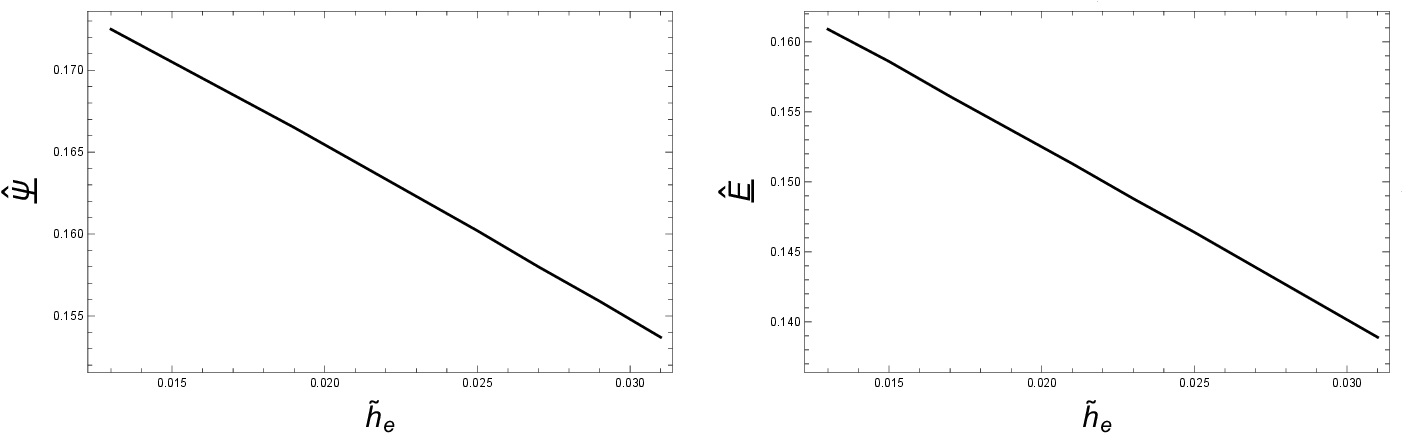}}\hfill
    \caption{Angular velocity $\underline{\hat{\psi}}$ (left panel) and Lyapunov exponent $\underline{\hat{E}}$ (right panel)
    versus quintessence parameter $\tilde{h}_{e}$ with a fixed parameters are $\check{\bar{M}} = 1$,
    $l_{y} = 2$, $\tilde{\sigma}_{e} = 0.5$, and, $\tilde{a}_{e} = 0.03$.}
    \label{fig:multi_graphs}\label{FIG.9}
\end{figure}

\section{Greybody Factors}

The discovery made by Hawking in $1975$ stated that BHs are not entirely dark
because, at this point of contact with the horizon of the BHs, they emit thermal
radiation because of the presence of a quantum effect therein \cite{11}. This
emission is exactly that of perfect blackbody radiation at the Hawking temperature,
even at its horizon. However, as the radiation propagates away in the curved
geometrical spacetime in the vicinity of the BH, part of it is reflected by the
effective potential barrier. This causes the observer at infinity to observe a
different spectrum, which is not a pure blackbody. 

This transformation is
quantified by the GFs, which are the frequency and angular momentum-dependent
outgoing Hawking probability of Hawking modes to propagate to infinity. Simply
stated, the GFs is a kind of a filter that warps the original
blackbody like radiation to produce what is alternatively referred to as
greybody radiation \cite{12,13}. There are a number of methods that can
be used to compute GFs, including those developed by
Fernando \cite{15}, Maldacena et al. \cite{14}, and others \cite{16,17,18,19,20}.

\subsection{GFs Analysis Based on the WKB Approach}

This subsection applies the 6th-order WKB approximation, which is a powerful
semi-analytical method in the BH perturbation theory and determines the GFs
of both the scalar and electromagnetic perturbations. We compute the transmission
and reflection coefficients on the basis of the conventional scattering boundary
conditions (incoming pure waves at infinity and no incoming waves at the horizon). 

We examine the effect of the parameter of the string cloud, the quantum correction
parameter, and the quintessence parameter on the greybody spectrum through the
6th-order WKB expansion. As shown in the below figures, the parameters have a
strong effect on altering the filtering behavior of the potential barrier, thus
changing the energy distribution and the intensity of the emitted Hawking
radiation. This discussion is a follow-up of our previous work in the QNMs, and it
offers more information on the thermodynamics and quantum dynamics of the BH.
\begin{equation}\label{11}
\hat{\underline{\phi}}=\check{\tilde{e}}^{-i\check{\varpi}{r}_{*}}+\hat{\bar{R}}\check{\tilde{e}}^{i\check{\varpi}{r}_{*}}~~as~~r_{*}\rightarrow+\infty,~~~~
\hat{\underline{\phi}}=\hat{\bar{T}}\check{\tilde{e}}^{-i\check{\varpi}{r}_{*}}~~as~~r_{*}\rightarrow-\infty,
\end{equation}
in which the transmission coefficient is denoted by the symbol of $\hat{\bar{T}}$, and the
reflection coefficient is denoted by the symbol of $\hat{\bar{R}}$. These coefficients
satisfy the conservation relation $|\hat{\bar{T}}|^{2}+|\hat{\bar{R}}|^{2} = 1$, which states
that the sum of the transmission and reflection probabilities must be equal to one.
\begin{equation}\label{12}
|\hat{\bar{T}}|^{2}=1-|\hat{\bar{R}}|^{2}=|\underline{\hat{N}}|^{2}.
\end{equation}
The following formula of WKB is used to determine the reflection coefficient $\hat{\bar{R}}$, which is denoted as
\begin{equation}\label{13}
\hat{\bar{R}}=\big(1+\check{\tilde{e}}^{-2i\pi\check{\kappa}}\big)^{-\frac{1}{2}}.
\end{equation}
To calculate the phase factor $\check{\kappa}$, using the following formula
\begin{equation}\label{14}
\check{\kappa}-i\frac{(\check{\varpi}^{2}-\bar{\check{V}}_{0})}{\sqrt{-2\bar{\check{V}}^{\prime\prime}_{0}}}-\sum_{b=2}^{6}\breve{\Psi}_{b}=0.
\end{equation}
Even though the 6th-order WKB approximation is very efficient in calculating GFs, it can be a little
less accurate at very low frequencies. The effective potential barrier in this regime is likely to
lead to almost total reflection and pushes the GFs to zero. Nonetheless, this weakness
does not have an impact on the determination of the energy emission rates. Since the WKB method is
reliable and applicable in both BH perturbation theory and greybody research, it is beyond
the scope of this work to review the foundations of the method in detail. The detailed literature
on the topic is addressed to the interested readers \cite{21,22}. 

The dynamics of the GFs of massless
scalar and electromagnetic perturbations in the presence of different string
clouds, quantum corrections, and quintessence parameters were carefully examined under the given
BH. Figure $\textbf{10}$ presents the following results. Each figure presents useful information on
the BH dynamics, which is a clear demonstration of how each parameter affects the absorption
probability using the GFs. In the following subsections, we discuss the effects
of the quintessence parameter, the multipole moment, the quantum correction parameter, and the
string cloud parameter on the GFs of massless scalar and electromagnetic perturbations
in a systematic manner.

\subsubsection{Behavior of GFs and Absorption Probabilities}

Figures \textbf{10} and \textbf{11} present the GFs of massless scalar and electromagnetic
perturbations of the QOS BH in the presence of quintessence and string
clouds. As can be easily observed in the plots, the GFs are highly dependent on
the model parameters. It is worth noting that the greater the multipole moment $l_{y}$, the smaller
the GFs. Such suppression is enhanced at larger values of $l_y$, which is explained by the
fact that higher angular momentum modes have a stronger centrifugal barrier. Therefore, the
chances of these modes tunneling through the effective potential barrier are lower. Equally, the
GFs have reduced systematically as the quantum correction parameter $\tilde{\sigma}_{e}$ is
increased. The behavior of this process is that the stronger the quantum corrections, the more
effective the backscattering, and it reduces the transmission probability of the Hawking modes. 

Based
on the tendencies, which could be noticed in the Figs. \textbf{10} and \textbf{11}, it is clear
that the increased values of the multipole moment, as well as the quantum correction parameter, lead
to the reduction of the GFs. This means that a smaller amount of radiation will
escape to infinity (i.e., the BH emits less Hawking radiation). This means that a
greater fraction of the radiation is reflected back or trapped, and so there is less
probability of absorption as viewed by an incoming wave, or more simply, reduced Hawking
radiation emission to distant observers.
In contrast to the multipole moment and quantum correction effects, the GFs have
shown an interesting opposite trend with the string cloud parameter
$\tilde{a}_{e}$ and the quintessence parameter $\tilde{h}_{e}$. 

As it is evident
in Figs. \textbf{10} and \textbf{11}, increasing values of these parameters lead to
a significant improvement of the GFs of the massless scalar and
electromagnetic perturbations. This increase in the GFs may be explained
by the fact that the height of the effective potential barrier gradually decreases
with the increase in the string cloud and quintessence parameters. The lower potential
barrier minimizes backscattering and enhances the easy penetration of the quantum modes
through the barrier. As a result, a greater proportion of the Hawking radiation is able
to escape to the faraway observers, causing increased rates of emissions. Correspondingly, the
probability of the incoming waves from infinity being absorbed also rises. The above outcomes
eloquently demonstrate how the clouds of surrounding strings and quintessential dark energy
can moderate the gravitational obstacle, which can, in turn, increase the transfer of
quantum fields in the vicinity of a QOS BH.
\begin{figure}[h!]
    \centering
   {\includegraphics[width=0.95\textwidth]{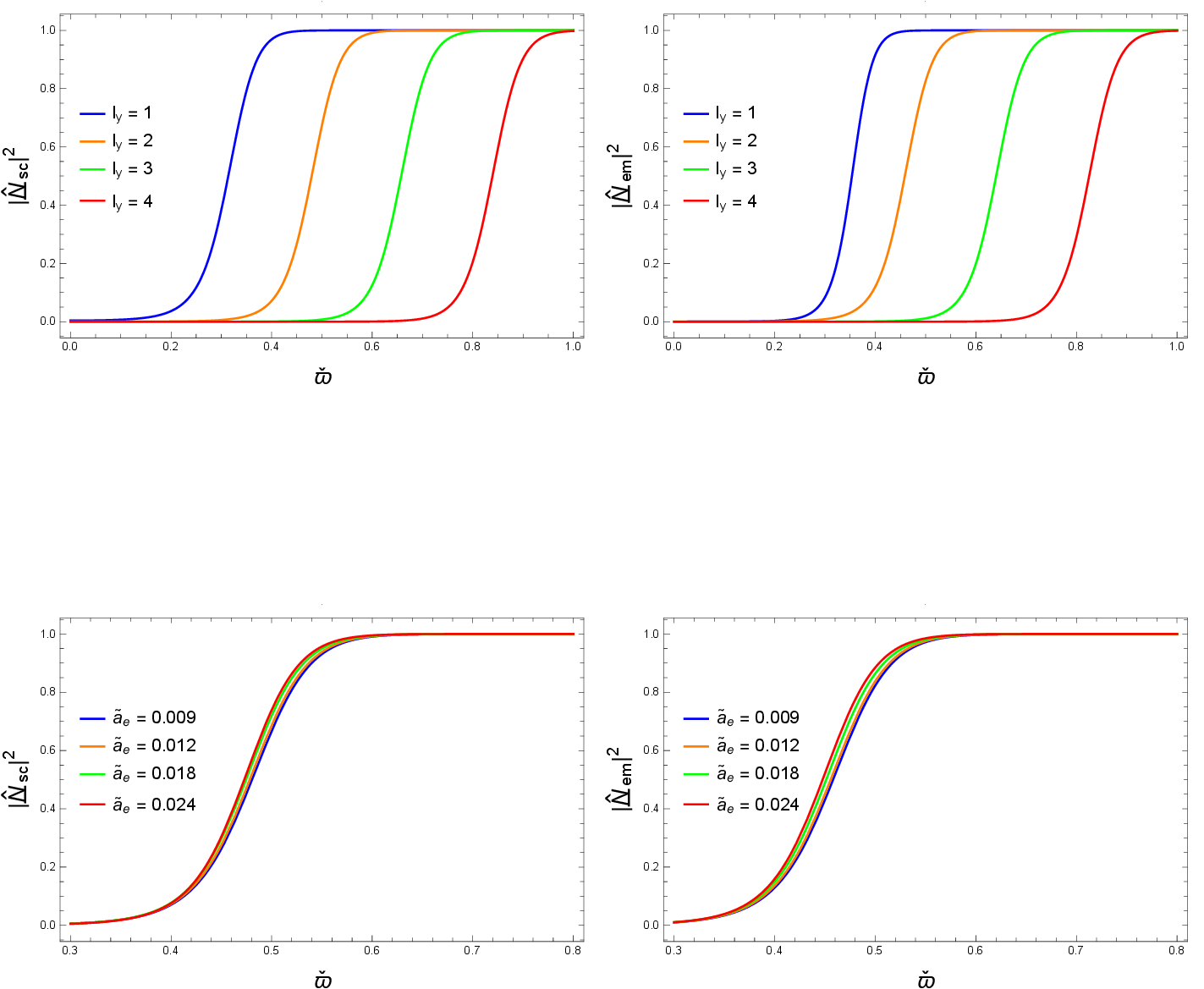}}\hfill
    \caption{Greybody factors of the massless scalar (left panel) and electromagnetic (right panel) perturbation at various values of the multipole moment $l_{y}$, and the string cloud parameter $\tilde{a}_{e}$, with constant parameters $\check{\bar{M}} = 1$, $\tilde{\sigma}_{e} = 0.9$, and, $\tilde{h}_{e} = 0.004$.}
    \label{fig:multi_graphs}\label{FIG.10}
\end{figure}

\begin{figure}[h!]
    \centering
   {\includegraphics[width=0.95\textwidth]{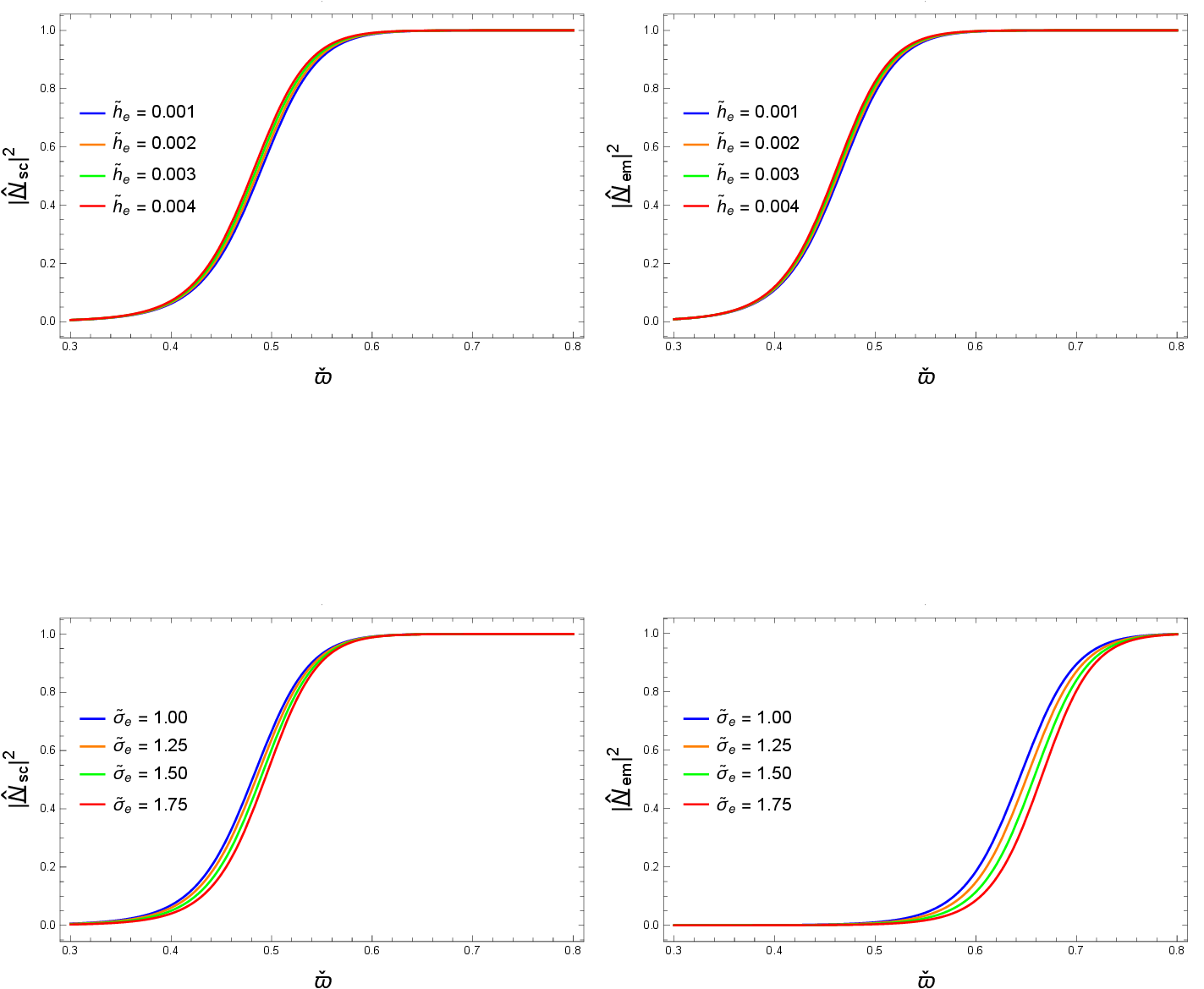}}\hfill
    \caption{Greybody factors of the massless scalar (left panel) and electromagnetic (right panel) perturbation at
    various values of the quintessence parameter $\tilde{h}_{e}$, and the quantum correction parameter $\tilde{\sigma}_{e}$, with fixed parameters are $\tilde{a}_{e} = 0.01$, $\check{\bar{M}} = 1$, and, $l_{y} = 2$.}
    \label{fig:multi_graphs}\label{FIG.11}
\end{figure}

\subsection{Study of Rigorous Bounds on Greybody Factors}

This section is based on a robust and beautiful strategy to obtain strict limits on the
GFs, mainly for scalar perturbations. It is noteworthy that both scalar and
electromagnetic perturbations can be seen to have tremendous similarities regarding their
GFs. The technique was initially proposed by Visser \cite{23} and subsequently
improved and expanded by Ngampitipan et al. \cite{24,25}, Boonserm et al. \cite{25,26,27,28,29,30}, and
many others. Using the structure of the preceding sections, we start with the Klein-Gordon equation
of a massless scalar field propagating in the spacetime of the QOS BH. We
have defined the effective potential, which appears in the radial wave equation, as follows
\begin{equation}\label{15}
\bar{\check{V}}_{eff}(r)=\hat{\bar{S}}(r) \Big[\frac{l_{y}(l_{y}+1)}{r^{2}}+\frac{1}{r}\frac{d\hat{\bar{S}}(r)}{dr}\Big].
\end{equation}
This effective potential determines the dynamics of perturbations of the scalar field in the BH
background, with $\hat{\bar{S}}(r)$ being the metric function and $l_{y}$ being the multipole
moment. With this useful potential, we obtain a strict lower bound on the GFs by
adhering to the beautiful strategy first formulated by Visser \cite{23} himself and later
improved by Boonserm et al. \cite{24,25}.  Here, $T_{\tilde{E}}$
represents the lower bounds of the transmission coefficient, which is defined as
\begin{equation}\label{16}
T_{\tilde{E}}\geq sech^{2}\bigg(\frac{1}{2\check{\varpi}}\int_{-\infty}^{+\infty}|\bar{\check{V}}_{eff}(r)| \frac{dr}{\hat{\bar{S}}(r)} \bigg),
\end{equation}
In this case, the frequency of the perturbation is denoted by $\check{\varpi}$. Boonserm et al. \cite{20}
made a beautiful change of the boundary conditions to introduce the equation of state parameter and the
coupling constant of the surrounding fields. The altered boundary conditions are presented below
 \begin{equation}\label{17}
\hat{G}\geq T_{\tilde{E}}= sech^{2}\bigg(\frac{1}{2\check{\varpi}}\int_{-\infty}^{+\infty}|\bar{\check{V}}_{eff}(r)| \frac{dr}{\hat{\bar{S}}(r)} \bigg)
=sech^{2} \bigg(\frac{F_{x}}{2\check{\varpi}}\bigg),
\end{equation}
with
\begin{equation}\label{18}
F_{x}= \int_{r_{H}}^{R_{H}}|\bar{\check{V}}_{eff}(r)| \frac{dr}{\hat{\bar{S}}(r)}
= \int_{r_{H}}^{R_{H}}\bigg|\frac{l_{y}(l_{y}+1)}{r^{2}}+\frac{1}{r}\frac{d\hat{\bar{S}}(r)}{dr}\bigg| dr.
\end{equation}
The above equations are used to obtain the effective potential between the cosmic horizon $R_{H}$ and
the event horizon $r_{H}$ of the black hole. Using this effective potential, we obtain strict constraints
on the GFs of the QOS BH in the presence of quintessential dark energy and
string clouds. As a result, the lower transmission coefficient bound on GFs has been determined.
\begin{eqnarray}\label{19}\notag
T_{\tilde{E}}=sech^{2}\bigg[\frac{1}{2\check{\varpi}}\bigg(\tilde{h}_{e} \left(\log \left(r_H\right)-\log \left(R_H\right)\right)-\frac{l_{y} (l_{y}+1) R_H+\check{\bar{M}}}{R_H^2}+\notag\\\frac{l_{y} (l_{y}+1)}{r_H}-
\frac{4 \tilde{\sigma}_{e}  \check{\bar{M}}^2}{5 r_H^5}+\frac{4 \tilde{\sigma}_{e}  \check{\bar{M}}^2}{5 R_H^5}+\frac{\check{\bar{M}}}{r_H^2}\bigg).
\end{eqnarray}
This term provides a strict lower limit of the GFs as a function of different BH
parameters. We demonstrate that these limits are directly controlled by the most important
parameters of the gravity model, namely the string cloud, quantum correction, and quintessence
parameters. These findings give interesting insights into the nature of Hawking radiation and
the complex interplay between the emitted radiation and the surrounding quintessential dark
energy and string clouds. The GFs of scalar and electromagnetic perturbations are
almost identical in behavior, so we simply and clearly consider only the scalar perturbations
in this subsection.

Figure \textbf{12} demonstrates the sensitivity of the rigorous lower bounds on massless scalar
perturbations on the different model parameters to the GFs. It is evident that the
various panels show the impact of these parameters on the probability of absorption, hence providing
useful information concerning the dynamical behavior of the BH with quintessential dark energy and string clouds.
\begin{figure}[h!]
 {\includegraphics[width=0.50\textwidth]{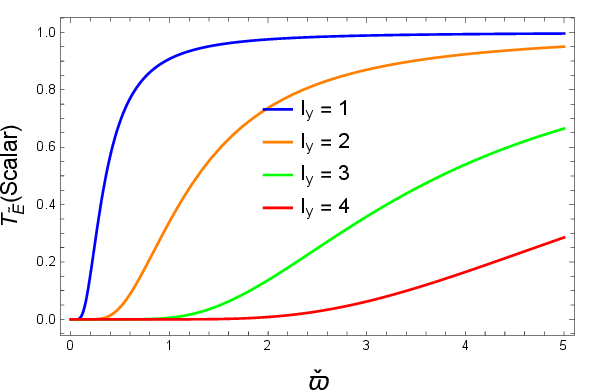}}
  {\includegraphics[width=0.50\textwidth]{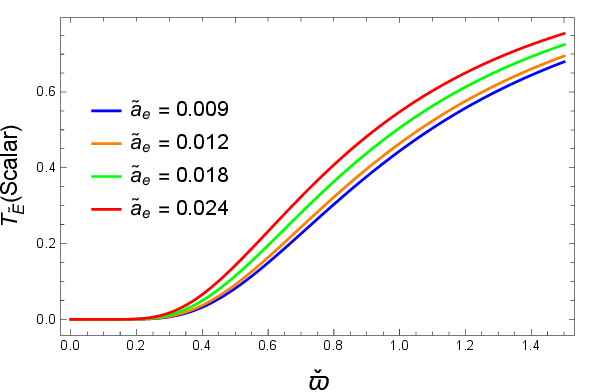}}
 {\includegraphics[width=0.50\textwidth]{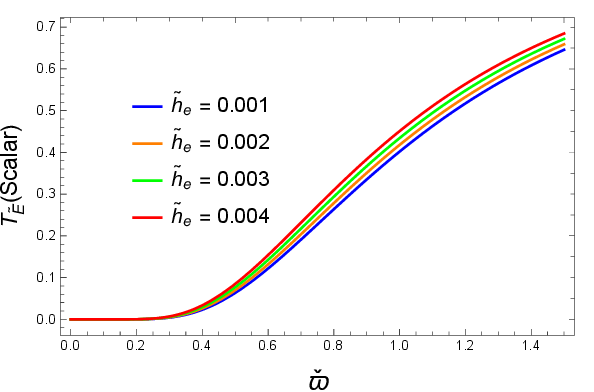}}
  {\includegraphics[width=0.50\textwidth]{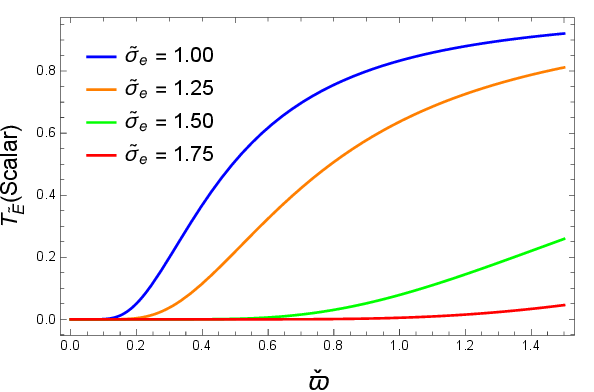}}
 \caption{Rigorous constraints on GFs are obtained at different values of the multipole moment $l_{y}$, string cloud
 parameter $\tilde{a}_{e}$, quintessence parameter $\tilde{h}_{e}$, and quantum correction parameter $\tilde{\sigma}_{e}$.}
    \label{fig:multi_graphs}\label{FIG.12}
\end{figure}

The strict lower bounds of the GFs of massless scalar perturbations are presented in
Fig. \textbf{12}. As can be seen from the panels, when the multipole moment and the
quantum correction parameter are increased, the lower bounds on the GFs
decrease noticeably. Conversely, a stronger enhancement of these bounds is evident
with an increase in the string cloud parameter and the quintessence parameter. This
act directly has physical consequences on Hawking radiation. The greater the lower
limit on the GFs, the greater the percentage of the emitted radiation
that is able to penetrate the potential barrier and be detected by distant
observers. 

Thus, raising the parameters of the string clouds and quintessence
enables greater emission of Hawking radiation, and raising the multipole moment
and quantum correction parameter inhibits the radiation emission. It is noteworthy
that the qualitative behavior of these strict lower bounds is the same as that of
the GFs themselves, which were estimated within the 6th-order WKB approximation in
the preceding subsection (illustrated in Figs. \textbf{10} and \textbf{11}).

\section{Conclusion}

We have explored the perturbation equations of two different types of fields with
different spins in the background of the QOS BH in the environment
of quintessential dark energy and string clouds. We have computed and plotted the effective
potential for these perturbations. Through our analysis, we find that the effective potential
peak values increase greatly with the increased multipole moment and the quantum correction
parameter, with the multipole moment having a remarkable effect on this increase.
Conversely, as the values of the string cloud parameter and the quintessence parameter
are increased, the height of the effective potential gradually decreases.

We have studied the QNMs and GFs of the QOS BH in the presence
of quintessential dark energy and string clouds, both through the 6th-order WKB approximation
and by paying attention to the scalar and electromagnetic perturbations, taking into
consideration the relevant and suitable boundary conditions. The findings of our work
indicate that the quintessence parameter, the string cloud parameter, and the multipole
moment are all important in determining both the QNMs and the GFs. Specifically, the multipole
moment causes the damping rates and real parts of the QNM frequencies to increase with the
multipole moment, which means that the restoring forces are stronger and the BH is more
stable to perturbations. In addition, we have also checked the quasinormal frequencies
with varying values of the string cloud parameter as well as the quintessence parameter. We
see that with an increase in the values of these parameters, the real and imaginary
components of the QNM frequencies decrease. This has been well demonstrated in the respective graphs.

In conclusion, all the computed QNM frequencies possess negative imaginary parts for
both scalar and electromagnetic perturbations. This affirms the dynamical stability
of the QOS BH to these perturbations. We find that as we raise
the quintessence parameter, and the string cloud parameter, both the real and the
magnitude of the imaginary part of the QNM frequencies decrease significantly. Physically, this
means that the BH vibrates with lower frequencies and slower oscillations, and the damping
is weaker. Thus, the ringout stage becomes longer but completely stable. These results
provide valuable information on the typical ringdown signatures and detectability of
gravitational waves of BHs in quintessential dark energy and string clouds. In addition, we
have determined the QNMs in the eikonal limit through the null geodesic method. An
elaborate graphical comparison between the results of the circular null geodesic
method in the eikonal regime and the 6th-order WKB approximation shows a great
deal of qualitative and quantitative agreement, which points to the validity of
the WKB method.

We have also calculated the GFs of the BH metric. We have plotted transmission
probability versus frequency for different model parameters. We find that the factors
of the greybody decrease with an increase in the values of the quantum correction
parameter and the multipole moment.
This decreases the transmission probability and makes the Hawking modes more strongly
backscattered by the effective potential barrier.
This means that a greater portion of the radiation is reflected back to the horizon, creating
a smaller Hawking flux when observed on a large scale.
On the other hand, an increase in the string cloud parameter and quintessence parameter
leads to the GFs growing significantly. This is beneficial as it increases
the tunneling probability of the potential barrier and minimizes backscattering and the
fraction of Hawking radiation that reaches remote observers. This causes the emitted
radiation to have greater intensity, which makes the BH look brighter in terms of its
Hawking emission. Lastly, we have obtained strict numerical lower limits on the greybody
contributions to scalar perturbations. It is remarkable that these bounds are controlled
by the same parameters and have precisely the same qualitative behavior as the GFs
themselves, giving consistent and independent validation of the trends observed.

These findings present new avenues of research in the future, especially on the phenomenological
implications of gravitational wave ringdown signals and Hawking radiation signatures of the
QOS BH in the presence of quintessential dark energy and string clouds.


\vspace{0.1cm}

\begin{thebibliography}{40}

\bibitem{1*} Voit, G. M., \& Shull, J. M. (1988). X-ray induced stellar mass loss near active galactic nuclei. Astrophysical Journal, Part 1 (ISSN 0004-637X), vol. 331, Aug. 1, 1988, p. 197-210., 331, 197-210.
\bibitem{2*} Blome, H. J., \& Mashhoon, B. (1984). Quasi-normal oscillations of a schwarzschild black hole. Physics Letters A, 100(5), 231-234.
\bibitem{3*} Vishveshwara, C. V. (1970). Scattering of gravitational radiation by a Schwarzschild black-hole. Nature, 227(5261), 936-938.
\bibitem{4*} Israel, W. (1967). Event horizons in static vacuum space-times. Physical review, 164(5), 1776.
\bibitem{5*} Carter, B. (1971). Axisymmetric black hole has only two degrees of freedom. Physical Review Letters, 26(6), 331.
\bibitem{6*} Modesto, L., \& Rachwał, L. (2017). Nonlocal quantum gravity: A review. International Journal of Modern Physics D, 26(11), 1730020.
\bibitem{7*} Ayon-Beato, E., \& Garcia, A. (1998). Regular black hole in general relativity coupled to nonlinear electrodynamics. Physical review letters, 80(23), 5056.
\bibitem{8*} Hayward, S. A. (2006). Formation and evaporation of nonsingular black holes. Physical review letters, 96(3), 031103.
\bibitem{9*} Barriola, M., \& Vilenkin, A. (1989). Gravitational field of a global monopole. Physical Review Letters, 63(4), 341.
\bibitem{10*} Harari, D., \& Lousto, C. (1990). Repulsive gravitational effects of global monopoles. Physical Review D, 42(8), 2626.
\bibitem{11*} Dadhich, N., Narayan, K., \& Yajnik, U. A. (1998). Schwarzschild black hole with global monopole charge. Pramana, 50(4), 307-314.
\bibitem{12*} Vilenkin, A. (1985). Cosmic strings and domain walls. Physics reports, 121(5), 263-315.
\bibitem{13*} Barriola, M., \& Vilenkin, A. (1989). Gravitational field of a global monopole. Physical Review Letters, 63(4), 341.
\bibitem{14*} Bronnikov, K. A., Meierovich, B. E., \& Podolyak, E. R. (2002). Global monopole in general relativity. Journal of Experimental and Theoretical Physics, 95(3), 392-403.
\bibitem{15*} Achucarro, A., Gregory, R., \& Kuijken, K. (1995). Abelian Higgs hair for black holes. arXiv preprint gr-qc/9505039.
\bibitem{16*} Babichev, E., \& Charmousis, C. (2014). Dressing a black hole with a time-dependent Galileon. Journal of High Energy Physics, 2014(8), 1-10.
\bibitem{17*} Singh, D. V., Upadhyay, S., Myrzakulov, Y., Myrzakulov, K., Singh, B., \& Kumar, M. (2025). Thermodynamic behavior and phase transitions of black holes with a cloud of strings and perfect fluid dark matter. Nuclear Physics B, 1016, 116915.
\bibitem{18*} Kokkotas, K. D., \& Schmidt, B. G. (1999). Quasi-normal modes of stars and black holes. Living Reviews in Relativity, 2(1), 2.
\bibitem{19*} Berti, E., Cardoso, V., \& Starinets, A. O. (2009). Quasinormal modes of black holes and black branes. Classical and Quantum Gravity, 26(16), 163001.
\bibitem{20*} Nollert, H. P. (1999). Quasinormal modes: the characteristicsound'of black holes and neutron stars. Classical and Quantum Gravity, 16(12), R159-R216.
\bibitem{21*} Ferrari, V., \& Gualtieri, L. (2008). Quasi-normal modes and gravitational wave astronomy. General Relativity and Gravitation, 40(5), 945-970.
\bibitem{22*} Cardoso, V., Franzin, E., \& Pani, P. (2016). Erratum: Is the gravitational-wave ringdown a probe of the event horizon?[phys. rev. lett. 116, 171101 (2016)]. Physical review letters, 117(8), 089902.
\bibitem{23*} Sotiriou, T. P., \& Faraoni, V. (2010). f (R) theories of gravity. Reviews of Modern Physics, 82(1), 451-497.
\bibitem{24*} Regge, T., \& Wheeler, J. A. (1957). Stability of a Schwarzschild singularity. Physical Review, 108(4), 1063.
\bibitem{25*} Chandrasekhar, S. (1998). The mathematical theory of black holes (Vol. 69). Oxford university press.
\bibitem{26*} Nollert, H. P. (1999). Quasinormal modes: the characteristicsound'of black holes and neutron stars. Classical and Quantum Gravity, 16(12), R159.
\bibitem{27*} Kokkotas, K. D., \& Schmidt, B. G. (1999). Quasi-normal modes of stars and black holes. Living Reviews in Relativity, 2(1), 2.
\bibitem{28*} Horowitz, G. T., \& Hubeny, V. E. (2000). Quasinormal modes of AdS black holes and the approach to thermal equilibrium. Physical Review D, 62(2), 024027.
\bibitem{29*} Cardoso, V., \& Lemos, J. P. (2001). Scalar, electromagnetic, and Weyl perturbations of BTZ black holes: Quasinormal modes. Physical Review D, 63(12), 124015.
\bibitem{30*} Moss, I. G., \& Norman, J. P. (2002). Gravitational quasinormal modes for anti-de Sitter black holes. Classical and Quantum Gravity, 19(8), 2323.
\bibitem{31*} Dreyer, O. (2003). Quasinormal modes, the area spectrum, and black hole entropy. Physical Review Letters, 90(8), 081301.
\bibitem{32*} Konoplya, R. A. (2002). Massive charged scalar field in a Reissner–Nordstrom black hole background: quasinormal ringing. Physics Letters B, 550(1-2), 117-120.
\bibitem{33*} Cardoso, V., \& Lemos, J. P. (2003). Quasinormal modes of the near extremal Schwarzschild–de Sitter black hole. Physical Review D, 67(8), 084020.
\bibitem{34*} Leaver, E. W. (1990). Quasinormal modes of Reissner-Nordström black holes. Physical Review D, 41(10), 2986.
\bibitem{35*} Kokkotas, K. D., \& Schutz, B. F. (1988). Black-hole normal modes: A WKB approach. III. The Reissner-Nordström black hole. Physical Review D, 37(12), 3378.
\bibitem{36*} Gonzalez, P. A., Övgün, A., Saavedra, J., \& Vásquez, Y. (2018). Hawking radiation and propagation of massive charged scalar field on a three-dimensional Gödel black hole. General Relativity and Gravitation, 50(6), 62.
\bibitem{37*} Yang, J., Zhang, C., \& Ma, Y. (2023). Shadow and stability of quantum-corrected black holes. The European Physical Journal C, 83(7), 619.
\bibitem{38*} Hawking, S. W. (1975). Particle creation by black holes. Communications in mathematical physics, 43(3), 199-220.


\bibitem{39*} Oshita, N. (2024). Greybody factors imprinted on black hole ringdowns: An alternative to superposed quasinormal modes. Physical Review D, 109(10), 104028.
\bibitem{40*} Konoplya, R. A.,  Zinhailo, A. F. (2019). Hawking radiation of non-Schwarzschild black holes in higher derivative gravity: a crucial role of grey body factors. Physical Review D, 99(10), 104060.
\bibitem{41*} Cardoso, V., Cavaglia, M.,  Gualtieri, L. (2006). Black hole particle emission in higher dimensional spacetimes. Physical review letters, 96(7), 071301.
\bibitem{42*} Dey, S., Chakrabarti, S. (2019). A note on electromagnetic and gravitational perturbations of the Bardeen de Sitter black hole: quasinormal modes and greybody factors. The European Physical Journal C, 79(6), 504.
\bibitem{43*} Boonserm, P., Ngampitipan, T., \& Wongjun, P. (2018). Greybody factor for black holes in dRGT massive gravity. The European Physical Journal C, 78(6), 492.
\bibitem{44*} Konoplya, R. A. (2003). Quasinormal behavior of the D-dimensional Schwarzschild black hole and the higher order WKB approach. Physical Review D, 68(2), 024018.
\bibitem{45*} Kokkotas, K. D., Konoplya, R. A.,  Zhidenko, A. (2011). Quasinormal modes, scattering, and Hawking radiation of<? format?> Kerr-Newman black holes in a magnetic field. Physical Review D Particles, Fields, Gravitation, and Cosmology, 83(2), 024031.
\bibitem{46*} Konoplya, R. A., Zinhailo, A. F., Stuchlik, Z. (2020). Quasinormal modes and Hawking radiation of black holes in cubic gravity. Physical Review D, 102(4), 044023.
\bibitem{47*} Konoplya, R. A.,  Zinhailo, A. F. (2020). Grey-body factors and Hawking radiation of black holes in 4D Einstein-Gauss-Bonnet gravity. Physics Letters B, 810, 135793.
\bibitem{48*} Li, Q., Ma, C., Zhang, Y., Lin, Z. W.,  Duan, P. F. (2022). Shadow, absorption and Hawking radiation of a Schwarzschild black hole surrounded by a cloud of strings in Rastall gravity. The European Physical Journal C, 82(7), 658.
    \bibitem{48**} Sajjad, W., \& Azam, M. (2026). Study of Kiselev black hole in quantum fluctuation modified gravity via quasinormal modes and greybody factors. Physica Scripta, 101(9), 095301.
        \bibitem{48***} Sajjad, W., Azam, M., Mushtaq, F., Al-Badawi, A., \& Jawad, A. (2026). Analysis of Kiselev black hole in $f(\tau,\mathrm{T})$ gravity through Quasinormal modes, greybody factors and Thermodynamic Quantities. Physics Letters A, 131625.
  \bibitem{49*} Visser, M. (1999). Some general bounds for one-dimensional scattering. Physical Review A, 59(1), 427.
  \bibitem{50*} Boonserm, P. (2009). Rigorous bounds on Transmission, Reflection, and Bogoliubov coefficients. arXiv preprint arXiv:0907.0045.
  \bibitem{51*} Boonserm, P., \& Visser, M. (2010). Reformulating the Schrödinger equation as a Shabat–Zakharov system. Journal of Mathematical Physics, 51(2).

%
%\bibitem{37*}
%\bibitem{37*}
%\bibitem{37*}
%\bibitem{37*}
%\bibitem{37*}
%\bibitem{37*}

\bibitem{1} Ahmed, F., Al-Badawi, A., \& Sakallı, İ. (2025). Quantum Oppenheimer-Snyder Black Hole with Quintessential Dark Energy and a String Clouds: Geodesics, Perturbative Dynamics, and Thermal Properties. arXiv preprint arXiv:2508.03202.
\bibitem{2} Regge, T., \& Wheeler, J. A. (1957). Stability of a Schwarzschild singularity. Physical Review, 108(4), 1063.
\bibitem{3} Medved, A. J. M., Martin, D., \& Visser, M. (2004). Dirty black holes: quasinormal modes. Classical and Quantum Gravity, 21(6), 1393.
\bibitem{4} Nomura, H., \& Tamaki, T. (2005). Continuous area spectrum of a regular black hole. Physical Review D—Particles, Fields, Gravitation, and Cosmology, 71(12), 124033.

\bibitem{5} Schutz, B. F., \& Will, C. M. (1985). Black hole normal modes: a semianalytic approach. The Astrophysical Journal, 291, L33-L36.
\bibitem{6} Iyer, S., \& Will, C. M. (1987). Black-hole normal modes: A WKB approach. I. Foundations and application of a higher-order WKB analysis of potential-barrier scattering. Physical Review D, 35(12), 3621.

\bibitem{7} Konoplya, R. A., Zhidenko, A., \& Zinhailo, A. F. (2019). Higher order WKB formula for quasinormal modes and grey-body factors: recipes for quick and accurate calculations. Classical and Quantum Gravity, 36(15), 155002.
\bibitem{8} Konoplya, R. A., \& Zhidenko, A. (2011). Quasinormal modes of black holes: From astrophysics to string theory. Reviews of Modern Physics, 83(3), 793-836.
\bibitem{9} Konoplya, R. A. (2003). Quasinormal behavior of the D-dimensional Schwarzschild black hole and the higher order WKB approach. Physical Review D, 68(2), 024018.
\bibitem{10NullQNMs} Cardoso, V., Miranda, A. S., Berti, E., Witek, H., \& Zanchin, V. T. (2009). Geodesic stability, Lyapunov exponents, and quasinormal modes. Physical Review D—Particles, Fields, Gravitation, and Cosmology, 79(6), 064016.
\bibitem{11} Hawking, S. W. (1975). Particle creation by black holes. Communications in mathematical physics, 43(3), 199-220.
\bibitem{12} Singleton, D., \& Wilburn, S. (2011). Hawking radiation, Unruh radiation, and the equivalence principle. Physical Review Letters, 107(8), 081102.
\bibitem{13} Akhmedova, V., Pilling, T., de Gill, A., \& Singleton, D. (2008). Temporal contribution to gravitational WKB-like calculations. Physics Letters B, 666(3), 269-271.
\bibitem{14} Maldacena, J., \& Strominger, A. (1997). Black hole greybody factors and D-brane spectroscopy. Physical Review D, 55(2), 861.
\bibitem{15} Fernando, S. (2005). Greybody factors of charged dilaton black holes in 2+ 1 dimensions. General Relativity and Gravitation, 37(3), 461-481.
\bibitem{16} Panotopoulos, G., \& Rincón, Á. (2017). Greybody factors for a nonminimally coupled scalar field in BTZ black hole background. Physics Letters B, 772, 523-528.
\bibitem{17} Ahmed, J., \& Saifullah, K. (2018). Greybody factor of a scalar field from Reissner–Nordström–de Sitter black hole. The European Physical Journal C, 78(4), 316.
\bibitem{18} Javed, W., Aqib, M., \& Övgün, A. (2022). Effect of the magnetic charge on weak deflection angle and greybody bound of the black hole in Einstein-Gauss-Bonnet gravity. Physics Letters B, 829, 137114.
\bibitem{19} Al-Badawi, A., Kanzi, S., \& Sakallı, İ. (2023). Fermionic and bosonic greybody factors as well as quasinormal modes for charged Taub NUT black holes. Annals of Physics, 452, 169294.
\bibitem{20} Boonserm, P., Ngampitipan, T., \& Wongjun, P. (2019). Greybody factor for black string in dRGT massive gravity. The European Physical Journal C, 79(4), 330.
\bibitem{21} Konoplya, R. A., Zhidenko, A., \& Zinhailo, A. F. (2019). Higher order WKB formula for quasinormal modes and grey-body factors: recipes for quick and accurate calculations. Classical and Quantum Gravity, 36(15), 155002.
\bibitem{22} Konoplya, R. A., \& Zhidenko, A. (2011). Quasinormal modes of black holes: From astrophysics to string theory. Reviews of Modern Physics, 83(3), 793-836.
  \bibitem{23} Visser, M. (1999). Some general bounds for one-dimensional scattering. Physical Review A, 59(1), 427.
 \bibitem{24} Ngampitipan, T.,  Boonserm, P. (2013). Bounding the greybody factors for non-rotating black holes. International Journal of Modern Physics D, 22(09), 1350058.
 \bibitem{25} Boonserm, P., Ngampitipan, T., \& Wongjun, P. (2018). Greybody factor for black holes in dRGT massive gravity. The European Physical Journal C, 78(6), 492.
 \bibitem{26} Boonserm, P.,  Visser, M. (2008). Bounding the greybody factors for Schwarzschild black holes. Physical Review D Particles, Fields, Gravitation, and Cosmology, 78(10), 101502.

 \bibitem{27} Gray, F.,  Visser, M. (2018). Greybody factors for Schwarzschild black holes: Path-ordered exponentials and product integrals. Universe, 4(9), 93.
\bibitem{28} Chowdhury, A.,  Banerjee, N. (2020). Greybody factor and sparsity of Hawking radiation from a charged spherical black hole with scalar hair. Physics Letters B, 805, 135417.
\bibitem{29} Miao, Y. G., Xu, Z. M. (2017). Hawking radiation of five-dimensional charged black holes with scalar fields. Physics Letters B, 772, 542-546.
\bibitem{30} Liu, Y. (2022). Hawking temperature and the bound on greybody factors in D= 4 double field theory. The European Physical Journal C, 82(11), 1054.




\end{thebibliography}
\end{document}